\documentclass[11pt,twoside,english]{article}
\usepackage[T1]{fontenc}
\usepackage[latin9]{inputenc}
\usepackage{geometry}
\usepackage{babel}
\usepackage{float}
\usepackage{amsthm}
\usepackage{amsmath}
\usepackage{amssymb}
\usepackage{graphicx}
\usepackage{setspace}
\makeatletter

\providecommand{\tabularnewline}{\\}

\numberwithin{equation}{section}
\numberwithin{figure}{section}

\@ifundefined{showcaptionsetup}{}{%
 \PassOptionsToPackage{caption=false}{subfig}}
\usepackage{subfig}
\makeatother

\begin{document}

\title{Theory of weakly nonlinear self sustained detonations}

\author{L. M. Faria$^{*}$, A. R. Kasimov%
\thanks{Applied Mathematics and Computational Science, King Abdullah University
of Science and Technology, Thuwal, Saudi Arabia%
} %
\thanks{Corresponding author, aslan.kasimov@kaust.edu.sa%
}, and R. R. Rosales %
\thanks{Massachusetts Institute of Technology, Department of Mathematics,
Cambridge, MA 02139, USA%
}}
\maketitle
\begin{abstract}
We propose a theory of weakly nonlinear multi-dimensional self sustained
detonations based on asymptotic analysis of the reactive compressible
Navier-Stokes equations. We show that these equations can be reduced
to a model consisting of a forced, unsteady, small disturbance, transonic
equation and a rate equation for the heat release. In one spatial
dimension, the model simplifies to a forced Burgers equation. Through
analysis, numerical calculations and comparison with the reactive
Euler equations, the model is demonstrated to capture such essential
dynamical characteristics of detonations as the steady-state structure,
the linear stability spectrum, the period-doubling sequence of bifurcations
and chaos in one-dimensional detonations and cellular structures in
multi-dimensional detonations. 
\end{abstract}

\section{Introduction}

The phenomenon of detonation was discovered in the late ninetheenth
century as a supersonic combustion wave that propagates around a thousand
times faster than an ordinary flame. The precise nature of detonation
was elucidated in the works of Mikhelson, Chapman and Jouguet at the
turn of the twentieth century and of Zel'dovich, von Neumann and D\"oring
in the 1940's (ZND theory, \cite{fickett2011detonation,Lee-2008}).
It was established that a detonation is a shock wave that compresses
and heats the reactive gas to temperatures sufficiently high to ignite
chemical reactions within a short distance from the shock. Gas expansion
caused by the heat release in these reactions creates pressure waves
that, on the one hand, support the propagation of the shock and, on
the other, accelerate the flow relative to the shock. A self-sustained
detonation is possible if such flow acceleration is sufficient to
make the flow reach sonic conditions relative to the shock. The presence
of a sonic state at some distance from the shock isolates the flow
between the shock and the sonic point from the influence of the conditions
downstream of the sonic locus, thus making the wave self sustained.
Most actual detonations are self sustained and therefore represent
an important class of detonations.

In ZND theory, a detonation is assumed to be one-dimensional and in
a steady-state in a Galilean frame moving with the wave.
The theory is successful in explaining the fundamental nature of a
detonation as a coupled shock-reaction zone system. However, in experiments
with gaseous detonations, it was observed as early as 1926 that detonations
tend to be unsteady and multi-dimensional with rather complex structures
and dynamics \cite{VMT63,fickett2011detonation,Lee-2008}. To explain
these dynamics, theoretical efforts focused on analysis of the stability
of ZND solutions began in the 1960's in the works of Zaidel, Pukhnachev,
Schelkin and, most comprehensively, Erpenbeck (see \cite{fickett2011detonation}
for an early literature review). The theory of linear stability of
gaseous detonations was found to be rather involved especially with
regard to specific numerical computations for real gaseous mixtures.
Much later, Lee and Stewart \cite{LeeStewart90} revisited the stability
problem and solved it for an idealized system by employing a relatively
straightforward normal-mode approach. Solving the linear stability
problem for real complex gaseous mixtures or for systems with losses
remains an open problem. 

Linear stability theory is successful in predicting, for idealized
systems, why and how ZND detonations are unstable to linear perturbations.
It is able to predict the neutral stability boundary and the most
unstable modes and thus the characteristic length scales of various
multi-dimensional structures seen in experiments. However, the predictive
power of linear stability theory is limited by its very nature as
a linear theory. Gaseous detonations are known to be highly nonlinear
and unsteady. The lead shock of such a detonation wave propagates
with strong oscillations in its velocity and with a highly non-uniform
flow behind the shock that involves additional shocks propagating
transversely to the lead shock with triple points forming at their
intersection. The triple points in complex detonation waves propagating
in tubes with walls covered with soot leave fish-scale traces called
detonation cells \cite{VMT63,Lee-2008,fickett2011detonation}, and
these detonations are termed cellular. 

Theoretical prediction of the origin and structure of cellular detonations
remains an outstanding and challenging open problem. However, some
progress has been achieved with use of the tools of asymptotic analysis
that allow for insight into the nature of the problem under various
limiting conditions. For example, large activation energy \cite{short1997multidimensional},
small heat-release \cite{short1999multi,clavin2002dynamics,clavin2012analytical},
slow time evolution and small shock curvature \cite{YaoStewart96,bdzil2012theory},
weak nonlinearity \cite{RosalesMajda:1983ly,BMR91,rosales1989diffraction},
strong overdrive \cite{clavin2012analytical} and other limiting assumptions
lead to relatively simple asymptotic models. These asymptotic theories
in general describe idealized systems and are thus also limited in
their predictive power. Nevertheless, they provide important information
about the mechanisms involved in the existence of particular qualitative
traits in the real phenomenon.

Two fundamental structural and dynamical properties of gaseous detonations
are first that they propagate in a galloping regime in narrow tubes,
as characterized by large amplitude pulsations in an essentially one-dimensional
wave, and second that the structure of detonation fronts in large
channels/tubes or open environments is cellular. From a theoretical
point of view, the basic problem is to describe and explain, at least
at a qualitative level, the origin and dynamics of these galloping
and cellular detonations. Importantly, such detonations have been
reproduced and extensively studied in numerical simulations of the
reactive\textbf{ }Navier-Stokes/Euler equations with simpified descriptions
of the chemical reactions and equations of state \cite{OranBoris}.
Thus, from a modeling standpoint, the use of these equations is frequently
appropriate. Some features of pulsating and cellular detonations have
been reproduced theoretically in \cite{StewartAslamYao96,clavin2012analytical}.
However, the following questions remain: 1) Can an asymptotic theory
predict the observed period-doubling sequence of bifurcations in one-dimensional
detonations and the self-sustained cellular structures in two-dimensional
detonations? 2) How quantitatively close are the asymptotic predictions
to the numerical results from the reactive Navier-Stokes/Euler equations?

Our aim here is to develop a theory that captures the linear and nonlinear
dynamics of self sustained detonations, especially with regard to
the bifurcation sequence seen in numerical simulations of one-dimensional
detonations and the cellular structures found in two-dimensional detonations.
Our theory is asymptotic and relies on a number of approximations,
namely small heat release, large activation energy, slow time evolution,
weak curvature and the Newtonian limit ($0<\gamma-1\ll1$, where $\gamma$
is the ratio of specific heats). We build on the theory developed
in \cite{rosales1989diffraction,RosalesMajda:1983ly} by employing,
as an additional approximation, the Newtonian limit (also used in
\cite{clavin2002dynamics} in one spatial dimension for Euler equations)
for two-dimensional detonations with retained dissipative effects.
The resulting system is a coupled set of three nonlinear partial differential
equations. When the dissipative terms are neglected, it is a hyperbolic
system for which we compute the traveling wave solutions analogous
to ZND waves, their multi-dimensional linear stability properties
and full nonlinear dynamics. We provide a quantitative comparison
with the results from the reactive Euler equations for all three asymptotic
predictions: the steady-state solutions, the linear stability spectrum
and the cellular structure. The analysis of the reduced system with
retained dissipative effects is postponed for future work.

As we show in Section \ref{sec:Weakly-nonlinear-approximation}, the
two-dimensional reactive Navier-Stokes equations reduce to a forced
version of the unsteady, small disturbance, transonic equations (UTSD)
given by 
\begin{eqnarray}
u_{\tau}+uu_{x}+v_{y} & = & -\frac{1}{2}\lambda_{x}+\nu u_{xx},\label{eq:asymptotic-model-u-1}\\
v_{x} & = & u_{y},\label{eq:asymptotic-model-v-1}\\
\lambda_{x} & = & -k(1-\lambda)\exp\left(\theta\left(\sqrt{q}u+q\lambda\right)\right),\label{eq:asymptotic-model-lambda-1}
\end{eqnarray}
where $u,v$ and $\lambda$ represent leading-order corrections to
the $x$ velocity, $y$ velocity and reaction progress variable. The
right-hand side of (\ref{eq:asymptotic-model-lambda-1}) is the leading-order
contribution of the reaction rate assumed for simplicity to follow
a single step Arrhenius kinetics. The parameters $\nu,k,\theta$ and
$q$ are the rescaled viscosity, pre-exponential factor, activation
energy and heat release, respectively. We find that this reduced asymptotic
model captures, at both the qualitative and quantitative levels, not
only the ZND structure, but also the linear stability spectrum, the
pulsating non-linear dynamics of one-dimensional detonations and the
cellular dynamics of two-dimensional detonations.

We also mention the attempts at understanding the nonlinear dynamics
of detonations via the use of qualitative models such as those introduced
by Fickett \cite{Fickett:1979ys}  and Majda \cite{Majda:1980zr}.
The idea behind these models is to produce simplified systems that,
although not derived from first principles, are capable of reproducing
the observed complexity of the solutions of reactive Euler equations
while considerably simplifying the analysis. The qualitative models
of Fickett and Majda are closely related to the weakly nonlinear theory
of detonations developed in \cite{RosalesMajda:1983ly}. Although
these models and the asymptotic theory of Rosales and Majda have enjoyed
some success in explaining certain features of different types of
traveling wave solutions of reactive Euler/Navier-Stokes equations
(i.e., weak and strong detonations), they were shown to lack the necessary
complexity needed to reproduce the dynamical properties of real detonations
with the rate functions used in prior work \cite{humpherys2013stability}.
However, simple \emph{ad hoc} modifications of these models can reproduce
much of the complexity of one-dimensional detonations as shown in
recent work \cite{Radulescu:2011fk,kasimov2013model,FariaKasimovRosales-SIAM2014}
and also explained in the present work.

The remainder of this paper is organized as follows. In Section \ref{sec:Governing-equations-of-reactive-flow},
we state the main governing equations together with the modeling assumptions
regarding the medium and the chemical reactions. In Section \ref{sec:Weakly-nonlinear-approximation},
we develop an asymptotic approximation of the governing equations
and obtain the weakly nonlinear reduced system. We then investigate
in Section \ref{sec:Traveling-wave-solutions-and-stability} the possible
traveling wave solutions of the asymptotic equations and their linear
stability properties. Both the traveling wave solutions and stability
spectrum of the asymptotic model are compared with their corresponding
results in the reactive Euler system. For the case with no dissipative
effects, predictions of the asymptotic model are calculated numerically
in Section \ref{sec:Nonlinear-dynamics}, and a quantitative comparison
with the predictions of the reactive Euler equations is presented
as well. Finally, in Section \ref{sec:Discussion-and-conclusions},
we discuss the results as well as point out some remaining open problems.

\section{The main governing equations of reactive flow\label{sec:Governing-equations-of-reactive-flow}}

We assume that the medium is described by the following system of
equations expressing, respectively, the laws of conservation of mass,
momentum and energy, and the chemical heat release: 
\begin{eqnarray}
\frac{\mathrm{D}\rho}{\mathrm{D}t}+\rho\nabla\cdot\mathbf{u} & = & 0,\label{eq:continuity}\\
\rho\frac{\mathrm{D}\mathbf{u}}{\mathrm{D}t} & = & \text{div}\left(\mathbf{T}\right),\label{eq:momentum}\\
\rho\frac{\mathrm{D}e}{\mathrm{D}t} & = & \mathbf{T}:\mathbf{D}-\text{div}\left(\mathbf{q}_{e}\right),\label{eq:energy}\\
\rho\frac{\mathrm{D}\Lambda}{\mathrm{D}t} & = & \rho\tilde{W}-\text{div}(\mathbf{q}_{\Lambda}).\label{eq:reaction}
\end{eqnarray}
Here, $\mathrm{D}/\mathrm{D}t=\partial/\partial t+\mathbf{u}\cdot\nabla$
is the material derivative, $\mathrm{v}=1/\rho$ is the specific volume,
$\rho$ is the density, $\mathbf{u}$ is the velocity, $\mathbf{T}$
is the Cauchy stress tensor, $\mathbf{D}=\left(\nabla\mathbf{u}+\nabla\mathbf{u}^{T}\right)/2$
is the deformation tensor, $\mathbf{T}:\mathbf{D}=\sum_{i,j}T_{ij}D_{ij}$
is the double contraction of tensors, $e=e_{i}-\tilde{Q}\Lambda$
is the total internal energy per unit mass, $\tilde{Q}$ is the heat
release per unit mass, $\mathbf{q}_{e}$ and $\mathbf{q}_{\Lambda}$
represent the flux of energy and species, respectively, $\tilde{W}(\Lambda,T)$
is the rate of reaction and $\Lambda$ is the reaction-progress variable
that changes from $\Lambda=0$ in the fresh mixture to $\Lambda=1$
in the fully burnt products. 

We make the following standard modeling assumptions (e.g., \cite{williams1985combustion}):
\begin{enumerate}
\item The fluid is Newtonian, with the Stokes assumption on the bulk viscosity,
so that \\ $\mathbf{T}=-\left(p+\frac{2}{3}\mu\text{div}\left(\mathbf{u}\right)\right)\mathbf{I}+2\mu\mathbf{D}$,
where $\mu$ is the dynamic viscosity, $p$ is the pressure and $\mathbf{I}$
is the identity tensor.
\item The species flux is given by Fick's law, $\mathbf{q}_{\Lambda}=\mbox{-\ensuremath{\rho}}d\nabla\Lambda$,
with $d$ denoting the diffusion coefficient. 
\item The energy flux has contributions from both the heat conduction (given
by Fourier's law) and the species diffusion, so that $\mathbf{q}_{e}=-\kappa\nabla T+\tilde{Q}\rho d\nabla\Lambda$,
where $\kappa$ is the heat diffusion coefficient. 
\item The medium is a perfect gas described by the ideal-gas equation of
state, $p=\rho RT$, with the internal energy given by $e_{i}=p\mathrm{v}/\left(\gamma-1\right)$,
where $R$ is the universal gas constant divided by the molecular
weight and $\gamma$ is the ratio of specific heats, assumed to be
constant.
\item For simplicity, we take the rate of reaction to be $\tilde{W}=\tilde{k}(1-\Lambda)\exp(-\tilde{E}/RT)$,
with the added ignition temperature assumption that $\tilde{W}=0$
for $T<T_{i}$ for some temperature, $T_{i}$. Here, $\tilde{k}$
is the rate constant and $\tilde{E}$ is the activation energy. More
general rate functions can be considered, in principle, as long as
appropriate sensitivity to temperature is preserved.
\end{enumerate}
With these assumptions, we can then rewrite (\ref{eq:continuity}-\ref{eq:reaction})
as 
\begin{eqnarray}
\frac{\mathrm{D}\rho}{\mathrm{D}t}+\rho\nabla\cdot\mathbf{u} & = & 0,\label{eq:ns-continuity}\\
\rho\frac{\mathrm{D}\mathbf{u}}{\mathrm{D}t} & = & \nabla\cdot\left(-\left(p+\frac{2}{3}\mu\text{div}\left(\mathbf{u}\right)\right)\mathbf{I}+2\mu\mathbf{D}\right),\label{eq:ns-momentum}\\
\rho\frac{\mathrm{D}e}{\mathrm{D}t} & = & -p\nabla\cdot\mathbf{u}-\frac{2}{3}\mu\left(\nabla\cdot\mathbf{u}\right)^{2}+\mu\left(\nabla\mathbf{u}:\nabla\mathbf{u}\right)+\mu\left(\nabla\mathbf{u}:\nabla\mathbf{u}^{T}\right)+\nabla\cdot\left(\kappa\nabla T-\tilde{Q}\rho d\nabla\Lambda\right),\qquad \label{eq:ns-energy}\\
\rho\frac{\mathrm{D}\Lambda}{\mathrm{D}t} & = & \mbox{\ensuremath{\rho}}\tilde{W}+\nabla\cdot(\mbox{\ensuremath{\rho}}d\nabla\Lambda).\label{eq:ns-reaction}
\end{eqnarray}
For the analysis that follows, it is convenient to use $e=e_{i}-\tilde{Q}\lambda=RT/(\gamma-1)-\tilde{Q}\Lambda$
to express the energy equation as 
\begin{eqnarray}
\rho\frac{\mathrm{D}T}{\mathrm{D}t}-\frac{\gamma-1}{R\gamma}\frac{\mathrm{D}p}{\mathrm{D}t} & = & \frac{\gamma-1}{R\gamma}\left(\tilde{Q}\rho\tilde{W}-\frac{2}{3}\mu\left(\nabla\cdot\mathbf{u}\right)^{2}+\mu\left(\nabla\mathbf{u}:\nabla\mathbf{u}\right)+\mu\left(\nabla\mathbf{u}:\nabla\mathbf{u}^{T}\right)+\nabla\cdot\left(d\nabla T\right)\right). \qquad \label{eq:ns-temperature}
\end{eqnarray}

We shall focus on the two-dimensional case for simplicity. Consider
a localized wave moving into an equilibrium, quiescent state and let
$\rho_{a}$, $p_{a}$, $T_{a}$ and $u_{a}=\sqrt{p_{a}/\rho_{a}}$
denote, respectively, the density, pressure, temperature and Newtonian
sound speed in the fresh mixture ahead of the wave. We rescale the
dependent variables with respect to this reference state. The independent
variables are rescaled as follows:
\begin{equation}
x=\frac{X-D_{0}t}{x_{0}},\quad y=\frac{Y}{y_{0}},\quad\tau=\frac{t}{t_{0}},\label{eq:x-y-t-rescaling-1-1}
\end{equation}
where $X$ and $Y$ are the original spatial variables and $D_{0}$
is a typical wave speed, which is to be determined in the process
of deriving the asymptotic model by requiring non-triviality of the
leading-order corrections. The length scales, $x_{0}$, $y_{0}$ and
the time scale, $t_{0}$, are chosen to reflect the appropriate physics
of weakly non-linear waves. We assume that $\epsilon=x_{0}/\left(u_{a}t_{0}\right)$
is small, which means that the spatial scale of interest in the $x$-direction,
which is\textbf{ }related to\textbf{ }the size of the reaction zone,
is small compared with the typical distance covered by acoustic waves
in time $t_{0}$. For the transverse dimension, we assume the scaling
$y_{0}=x_{0}/\sqrt{\epsilon}$ . This follows from the fact that,
along a weakly curved front, a distance $\epsilon$ in the normal
direction corresponds to a distance $O(\sqrt{\epsilon})$ in the transverse
direction.

Several dimensionless groups appear upon rescaling of the governing
equations. We define the Reynolds, Pradtl and Lewis numbers, respectively,
as follows:

\begin{equation}
\mathrm{Re}=\frac{\rho_{a}u_{a}x_{0}}{\mu},\quad\mathrm{Pr}=\frac{c_{p}\mu}{\kappa},\quad\mathrm{Le}=\frac{\kappa}{\rho_{a}c_{p}d},
\end{equation}
where $c_{p}=\gamma R/\left(\gamma-1\right)$. Writing $\mathbf{u}=\left(U,V\right)^{T}$,
it is convenient to introduce the differential operator: 
\begin{equation}
L=\partial_{\tau}+\frac{1}{\epsilon}(U-D_{0})\partial_{x}+\frac{1}{\sqrt{\epsilon}}V\partial_{y}.
\end{equation}
Introducing the dimensionless parameters,
\begin{equation}
Q=\frac{\tilde{Q}}{RT_{a}},\quad E=\frac{\tilde{E}}{RT_{a}},\quad K=t_{0}\tilde{k}\exp\left(-E\right),
\end{equation}
the non-dimensional governing equations become:

\begin{eqnarray}
L[\rho]+\rho\left(\frac{1}{\epsilon}U_{x}+\frac{1}{\sqrt{\epsilon}}V_{y}\right) & = & 0,\label{eq:ns-continuity-dimensionless}\\
\rho L[U]+\frac{1}{\epsilon}p_{x} & = & \frac{1}{3\epsilon\mathrm{Re}}\left(U_{xx}+\sqrt{\epsilon}V_{xy}\right)+\frac{1}{\mathrm{\epsilon Re}}\left(U_{xx}+\epsilon U_{yy}\right),\label{eq:ns-x-momentum-dimensionless}\\
\rho L[V]+\frac{1}{\sqrt{\epsilon}}p_{y} & = & \frac{1}{3\epsilon\mathrm{Re}}\left(\sqrt{\epsilon}U_{xy}+\epsilon V_{yy}\right)+\frac{1}{\epsilon\mathrm{Re}}\left(V_{xx}+\epsilon V_{yy}\right),\label{eq:ns-y-momentum-dimensionless}\\
\rho L[T]-\frac{\left(\gamma-1\right)}{\gamma}L[p] & = & \frac{\gamma-1}{\gamma}\left(Q\rho W-\frac{2}{3\epsilon\mathrm{Re}}\left(U_{x}+\sqrt{\epsilon}V_{y}\right)^{2}+\frac{1}{\epsilon\mathrm{Re}}\left(U_{x}^{2}+\epsilon U_{y}^{2}+V_{x}^{2}+\epsilon V_{y}^{2}\right)\right)\nonumber \\
 &  & +\frac{\gamma-1}{\gamma}\frac{1}{\mathrm{\epsilon Re}}\left(U_{x}^{2}+\sqrt{\epsilon}U_{y}V_{x}+\sqrt{\epsilon}V_{x}U_{y}+\epsilon V_{y}^{2}\right)+\frac{1}{\mathrm{\epsilon Re}\mathrm{Pr}}\left(T_{xx}+\epsilon T_{yy}\right),\quad \qquad\label{eq:ns-temperature-dimensionless}\\
\rho L[\Lambda] & = & \rho W+\frac{1}{\mathrm{\epsilon Re}\mathrm{Pr}\mathrm{Le}}\left(\left(\rho\Lambda_{x}\right)_{x}+\epsilon\left(\rho\Lambda_{y}\right)_{y}\right),\label{eq:ns-reaction-dimensionless}
\end{eqnarray}
where $W$ is defined as 
\begin{equation}
W=K(1-\Lambda)\exp\left[E\left(1-\frac{1}{T}\right)\right].\label{eq:W-dimensionless}
\end{equation}

\section{Weakly nonlinear approximation\label{sec:Weakly-nonlinear-approximation}}

We develop an asymptotic simplification of the above general formulation
by considering a weakly nonlinear detonation wave, for which we assume
that the heat release is small, the activation energy is large and
the Newtonian limit, $0<\gamma-1\ll1$, holds.\textbf{ }To be precise,
we start from (\ref{eq:ns-continuity-dimensionless}-\ref{eq:ns-reaction-dimensionless})
and make the following assumptions:
\begin{enumerate}
\item $K=k/\epsilon$, $k=O(1)$. This assumption is chosen to ensure that
the reaction rate affects the leading order expansion of $\Lambda$.
Since $K=t_{0}\tilde{k}$, this assumption implies that the characteristic
time scale, $t_{0}$, of weakly nonlinear detonations is large compared
to the collision time, $1/\tilde{k}$, i.e., $t_{0}\sim(1/\tilde{k})/\epsilon$.
\item $\left(\gamma-1\right)Q/\gamma=\epsilon^{2}q$, $q=O\left(1\right)$.
This assumption implies that the heat release does not play a role
at the linear level. It does not mean that the chemistry is unimportant,
but that the heat release must have the appropriate size to balance
the nonlinear effects. The extra factor of $\left(\gamma-1\right)/\gamma$
in front of $Q$ arises naturally in the governing equations (see
(\ref{eq:ns-temperature-dimensionless})) and is retained in the definition
of $q$. With the further assumption below of small $\gamma-1$, this
implies that $Q$ is $O\left(\epsilon\right)$.
\item $E=\theta/\epsilon^{2},$ $\theta=O\left(1\right)$. This ensures
that small temperature deviations -- which are $O\left(\epsilon^{2}\right)$
for weak shocks in the Newtonian limit  -- have an $O\left(1\right)$
influence on the reaction rate.
\item $\gamma-1=\gamma_{1}\epsilon$, $\gamma_{1}=O\left(1\right)$. This
assumption is needed to balance the temperature fluctuations with
both the acoustics and chemistry at the same order. Without this assumption,
the leading order corrections for density, velocity and temperature
all behave the same way, as in a weakly nonlinear inert shock \cite{hunter1995asymptotic,rosales1991introduction}.
As we show later, having a temperature profile that is different from
density/velocity profiles is crucial, as it allows the model derived
here to incorporate the dynamical instabilities of detonation waves.
\item $\mathrm{Le}$ and $\mathrm{Pr}$ are $O\left(1\right)$, while $\mathrm{Re}$
is $O(1/\epsilon)$. Other scalings that highlight different transport
effects are of course possible, but they are not considered in this
work. 
\end{enumerate}

To understand some of the intuition behind the asymptotic ordering
above, recall the following well-known fact for weak shocks \cite{Whitham74}.
If the shock strength is measured by the relative jump in pressure
across the shock, $\Delta p=\left(p_{s}-p_{a}\right)/p_{a}$ (subscripts
$s$ and $a$ denoting post- and pre-shock states, respectively),
then the shock Mach number is $\mathrm{M}=1+\left[\left(\gamma+1\right)/(4\gamma)\right]\,\Delta p+O\left(\left(\Delta p\right)^{2}\right)$
and the shock temperature is $T_{s}/T_{a}=1+\left[\left(\gamma-1\right)/\gamma\right]\,\Delta p+O\left(\left(\Delta p\right)^{2}\right)$.
Therefore, for weak shocks, with $\mathrm{M}-1=O\left(\epsilon\right)$,
in the Newtonian limit, $\gamma-1=O\left(\epsilon\right)$, the leading-order
temperature correction is $O\left(\epsilon^{2}\right)$. That is,
all variables have an $O\left(\epsilon\right)$ jump across the shock,
but the temperature jump is smaller, only $O\left(\epsilon^{2}\right)$.
In the chosen asymptotic approximation, we therefore expect similar
temperature behavior in the reaction zone as well, at least with inviscid
detonations. \\
\indent Now, we assume the following expansions in the reaction zone:
\begin{align}
\rho & =1+\rho_{1}\epsilon+\rho_{2}\epsilon^{3/2}+\rho_{3}\epsilon^{2}+o(\epsilon^{2}),\\
T & =1+T_{1}\epsilon+T_{2}\epsilon^{3/2}+T_{3}\epsilon^{2}+o(\epsilon^{2}),\\
p & =1+p_{1}\epsilon+p_{2}\epsilon^{3/2}+p_{3}\epsilon^{2}+o(\epsilon^{2}),\\
\boldsymbol{u} & =\mathbf{u}_{1}\epsilon+\mathbf{u}_{2}\epsilon^{3/2}+\mathbf{u}_{3}\epsilon^{2}+o(\epsilon^{2}),\\
\Lambda & =\lambda+o(\epsilon).
\end{align}
The fractional powers appear here because we aim at capturing weak
curvature effects in the detonation front. These effects induce a
flow velocity transverse to the front, which is $O(\sqrt{\epsilon})$
smaller than the longitudinal velocity. Expansions of this type are
standard for waves incorporating the weak-curvature effect (e.g.,
\cite{keller1978rays}).

Expanding $p=\rho T$, we find that $p_{1}=\rho_{1}+T_{1}$, $p_{2}=\rho_{2}+T_{2}$
and $p_{3}=\rho_{3}+T_{3}+\rho_{1}T_{1}$. Using these relations to
eliminate pressure perturbations, (\ref{eq:ns-continuity-dimensionless}-\ref{eq:ns-reaction-dimensionless})
yield 
\begin{eqnarray}
\left(-D_{0}\rho_{1x}+U_{1x}\right)+\sqrt{\epsilon}\left(-D_{0}\rho_{2x}+U_{2x}+V_{1y}\right)+\nonumber \\
\epsilon\left(\rho_{1\tau}-D_{0}\rho_{3x}+U_{1}\rho_{1x}+U_{3x}+\rho_{1}U_{1x}+V_{2y}\right) & = & o(\epsilon),\\
\left(-D_{0}U_{1x}+T_{1x}+\rho_{1x}\right)+\sqrt{\epsilon}\left(-D_{0}U_{2x}+T_{2x}+\rho_{2x}\right)+\nonumber \\
\epsilon\left(U_{1\tau}-D_{0}U_{3x}+U_{1}U_{1x}+T_{1}\rho_{1x}+T_{3x}+\rho_{3x}-\rho_{1}\rho_{1x}\right) & = & \frac{4}{3}\frac{1}{\mathrm{Re}}\left(U_{1}\right)_{xx}+o(\epsilon),\\
\left(-D_{0}V_{1x}\right)+\sqrt{\epsilon}\left(-D_{0}V_{2x}+T_{1y}+\rho_{1y}\right)+\nonumber \\
\epsilon\left(V_{1\tau}-D_{0}V_{3x}+U_{1}V_{1x}+T_{2y}+\rho_{2y}\right) & = & \frac{1}{\mathrm{Re}}\left(V_{1}\right)_{xx}+o(\epsilon),\\
\left(-D_{0}T_{1x}\right)+\sqrt{\epsilon}\left(-D_{0}T_{2x}\right)+\nonumber \\
\epsilon\left(T_{1\tau}-D_{0}T_{3x}+U_{1}T_{1x}+\frac{\gamma_{1}}{\gamma}D_{0}\left(\rho_{1x}+T_{1x}\right)\right) & = & \epsilon q\omega+o(\epsilon),\\
-D_{0}\lambda_{x} & = & \omega+o(1).
\end{eqnarray}
Because of the weak heat release assumption, we need to expand only
the reaction progress variable and the reaction rate to the leading
order. As we shall see later, the leading-order corrections to temperature
are of order $\epsilon^{2}$, i.e., $T_{1}=T_{2}=0$, and therefore
the leading-order reaction rate is given by 
\begin{equation}
\omega=k(1-\lambda)\exp\left(\theta T_{3}\right).\label{eq:omega}
\end{equation}

Balancing $O(1)$ terms, we find that 
\begin{equation}
\left[\begin{array}{rrrr}
-D_{0} & 1 & 0 & 0\\
1 & -D_{0} & 0 & 1\\
0 & 0 & -1 & 0\\
0 & 0 & 0 & -1
\end{array}\right]\left[\begin{array}{r}
\rho\\
U\\
V\\
T
\end{array}\right]_{1x}=\left[\begin{array}{r}
0\\
0\\
0\\
0
\end{array}\right].\label{eq:O1-system}
\end{equation}
This homogeneous system has non-trivial solutions if and only if the
coefficient matrix (denoted by $\mathbf{A}$ from now on) is singular.
Therefore, we must have $D_{0}=\pm1$. We focus on a right-going wave
for which $D_{0}=1$. 

After substituting $D_{0}=1,$ we integrate the first equation in
(\ref{eq:O1-system}) to obtain $U_{1}=u(x,y,\tau)+\bar{U}(y,\tau),\ \rho_{1}=u(x,y,\tau)+\bar{\rho}\left(y,\tau\right)$
for some, so far arbitrary, functions $\bar{U}\left(y,t\right)$ and
$\bar{\rho}\left(y,t\right)$.\textbf{ }In order for our asymptotic
expansions to hold as we approach the upstream state, which is assumed
to be quiescent and uniform at all times, we require $\rho_{1}=U_{1}\equiv0$
as $x\to\infty$ (upstream of the wave). As a consequence, it follows
that $\bar{U}(y,t)=\bar{\rho}\left(y,t\right)\equiv0$. Had we considered
the case where the wave moves into a non-uniform background, the functions
$\bar{\rho}$ and $\bar{U}$ would provide the freedom needed to match
the asymptotic expansion in the wave zone to the flow ahead (see,
e.g., \cite{RosalesMajda:1983ly}). A similar reasoning can be used
to deduce from the third and fourth equation in (\ref{eq:O1-system})
that $T_{1}=V_{1}\equiv0$. 

At $O(\epsilon^{1/2})$, we obtain 
\begin{equation}
\left[\begin{array}{rrrr}
-1 & 1 & 0 & 0\\
1 & -1 & 0 & 1\\
0 & 0 & -1 & 0\\
0 & 0 & 0 & -1
\end{array}\right]\left[\begin{array}{r}
\rho\\
U\\
V\\
T
\end{array}\right]_{2x}=\left[\begin{array}{r}
0\\
0\\
-u_{y}\\
0
\end{array}\right],\label{eq:O-half-system}
\end{equation}
with the same matrix of coefficients, \textbf{$\mathbf{A}$}, as before.
Since $\mathbf{A}$ is singular, this system has solutions if and
only if the right-hand side is orthogonal to $\mathbf{l}_{0}=\left[1\,1\,0\,1\right]$,
which spans the left null-space of $\mathbf{A}$. Clearly, this condition
is always satisfied here. Note that from the third and fourth equation
in \ref{eq:O-half-system}, we obtain that $V_{2x}=u_{y}$ and $T_{2}=0$,
i.e., letting 
\begin{equation}
v=V_{2}\label{eq:v-def}
\end{equation}
such that 
\begin{equation}
v_{x}=u_{y}.\label{eq:vx-uy}
\end{equation}

Finally, at $O\left(\epsilon\right)$, we obtain 
\begin{equation}
\left[\begin{array}{rrrr}
-1 & 1 & 0 & 0\\
1 & -1 & 0 & 1\\
0 & 0 & -1 & 0\\
0 & 0 & 0 & -1
\end{array}\right]\left[\begin{array}{r}
\rho\\
U\\
V\\
T
\end{array}\right]_{3x}=\left[\begin{array}{r}
-u_{\tau}-2uu_{x}-v_{y}\\
-u_{\tau}+\frac{4}{3}\frac{1}{\mathrm{\epsilon Re}}u_{xx}\\
-\rho_{2y}\\
q\omega-\frac{\gamma_{1}}{\gamma}u_{x}
\end{array}\right].\label{eq:O-epsilon-system}
\end{equation}
Notice that since $\mathrm{Re}=O(1/\epsilon)$, all terms in (\ref{eq:O-epsilon-system})
are $O\left(1\right)$. The solvability condition for this system
is that the right-hand side be orthogonal to $\mathbf{l}_{0}=\left[1\,1\,0\,1\right]$,
which yields: 
\begin{eqnarray}
2u_{\tau}+2uu_{x}+v_{y} & = & q\omega-\frac{\gamma_{1}}{\gamma}u_{x}+\frac{4}{3}\frac{1}{\mathrm{\epsilon Re}}u_{xx}.
\end{eqnarray}
This equation together with
\begin{eqnarray}
v_{x} & = & u_{y},\\
\lambda_{x} & = & -\omega,
\end{eqnarray}
forms a closed system of equations. The temperature dependence in
the leading order rate function, (\ref{eq:omega}), can be eliminated
by integrating the last equation in (\ref{eq:O-epsilon-system}),
\begin{equation}
T_{3}=\frac{\gamma_{1}}{\gamma}u+q\lambda,\label{eq:T3}
\end{equation}
giving 
\begin{equation}
\omega=k(1-\lambda)\exp\left(\theta\left(\frac{\gamma_{1}}{\gamma}u+q\lambda\right)\right).\label{eq:omega-new}
\end{equation}
Therefore, we obtain 
\begin{eqnarray}
2u_{\tau}+2uu_{x}+\frac{\gamma_{1}}{\gamma}u_{x}+v_{y} & = & -q\lambda_{x}+\frac{4}{3}\frac{1}{\epsilon\mathrm{Re}}u_{xx},\\
v_{x} & = & u_{y},\\
\lambda_{x} & = & -k(1-\lambda)\exp\left(\theta\left(\frac{\gamma_{1}}{\gamma}u+q\lambda\right)\right).
\end{eqnarray}

As can be seen from (\ref{eq:O-epsilon-system}), species and heat
diffusion play no role in the asymptotic equations up to $O\left(\epsilon\right)$.
Had we considered different asymptotic orderings for the Lewis and
Prandtl numbers, these effects would introduce diffusive terms in
the energy and the reaction-rate equations.\textbf{ }We consider no
such effects in this work and hence the only effective diffusion in
the asymptotic model comes from the viscous dissipation.

It is convenient to further rescale the variables as: 
\begin{equation}
x\mapsto x-\gamma_{1}/\left(2\gamma\right)\tau,\ y\mapsto2^{-1/2}q^{-1/4}y,\ \tau\mapsto q^{-1/2}\tau,\ u\mapsto q^{1/2}u,\ v\mapsto2^{1/2}q^{3/4}v,
\end{equation}
where the scale for $u$ is chosen so that the traveling wave solution
found later in Section \ref{sub:Travelling-wave-solutions} has speed
$1$. We also choose $\epsilon=\left(\gamma-1\right)/\gamma$, which
means that $\gamma_{1}=\gamma$. Other choices of $\epsilon$ are
possible as long as $\epsilon$ and $\gamma-1$ are of the same asymptotic
order when $\epsilon\to0$. These choices, although equivalent in
the limit $\epsilon\rightarrow0$, will have some effect on the quantitative
numerical predictions for finite $\epsilon$. With this rescaling
and the choice of $\epsilon$, we obtain our final asymptotic system
of equations of weakly nonlinear detonation: 
\begin{eqnarray}
u_{\tau}+uu_{x}+v_{y} & = & -\frac{1}{2}\lambda_{x}+\mbox{\ensuremath{\nu}}u_{xx},\label{eq:asymptotic-model-u}\\
v_{x} & = & u_{y},\label{eq:asymptotic-model-v}\\
\lambda_{x} & = & -k(1-\lambda)\exp\left[\theta\left(\sqrt{q}u+q\lambda\right)\right],\label{eq:asymptotic-model-lambda}
\end{eqnarray}
where $\nu=2/\left(3\sqrt{q}\epsilon\mathrm{Re}\right)$ is the dimensionless
viscosity coefficient.

In the inviscid case, (\ref{eq:asymptotic-model-u}-\ref{eq:asymptotic-model-lambda})
must be supplemented by the appropriate jump conditions across shocks.
If the shock locus is defined by $\phi(x,y,\tau)=x-s(y,\tau)=0$ and
$[z]$ denotes the jump of $z$ across the shock, the Rankine-Hugoniot
conditions are: 
\begin{eqnarray}
s_{\tau}\left[u\right]-\frac{1}{2}\left[u^{2}\right]+s_{y}\left[v\right] & = & 0,\label{eq:asymptotic-jump-condition1}\\
s_{y}\left[u\right]+\left[v\right] & = & 0,\label{eq:asymptotic-jump-condition2}\\
\left[\lambda\right] & = & 0.\label{eq:asymptotic-jump-condition3}
\end{eqnarray}
These equations follow from the conservation form of (\ref{eq:asymptotic-model-u}-\ref{eq:asymptotic-model-lambda}). 

Implicit in the definition (\ref{eq:W-dimensionless}) of the rate
of reaction, $W$, lies the ignition temperature assumption, such
that $W\equiv0$ for $T<T_{i}/T_{a}$. Thus, the leading order reaction
rate, $\omega$, given by (\ref{eq:omega}) also satisfies $\omega\equiv0$
for $T_{3}<\left(T_{i}/T_{a}-1\right)/\epsilon^{2}$ or, equivalently,
for $u+q\lambda<\left(T_{i}/T_{a}-1\right)/\epsilon^{2}$. In order
to prevent reactions from occurring in the ambient state, a reasonable
constraint on the ignition temperature is that it be larger than the
ambient temperature, $T_{a}$. Furthermore, for the ZND solution to
exist, it is necessary that the ignition temperature, $T_{i}$, be
smaller than the temperature at the von Neumann state of the ZND solution.
In the weakly nonlinear regime considered in this paper, we must therefore
have $0<T_{i}-T_{a}=O\left(\epsilon^{2}\right)$. This should be viewed
as a modeling assumption about the chemical kinetics. Notice that
in the inviscid case, the ignition temperature, if taken to be $T_{i}\gtrapprox T_{a}$,
has no effect on the ZND profiles and their stability properties.
Simply, it states that the reactions occur only after the shock and
therefore, $\omega=0$ ahead of the wave. When considering viscous
effects, however, the traveling wave profiles and their dynamical
evolution will depend on the ignition temperature and $T_{i}$ should
therefore be considered as another parameter that affects the properties
of the solutions. 

For the convenience of the reader, in Table \ref{tab:Summary-of-parameters},
we collect the relations between various dimensionless and dimensional quantities. 

\begin{table}[h]
\begin{centering}
\begin{tabular}{|c|c|c|}
\hline 
Dimensional  & Dimensionless  & \textbf{ }Physical meaning\tabularnewline
\hline 
\hline 
$\tilde{Q}$ & $Q=\tilde{Q}/RT_{a}=\epsilon q$ & Heat release\tabularnewline
\hline 
$\tilde{E}$ & $E=\tilde{E}/RT_{a}=\theta/\epsilon^{2}$ & Activation energy\tabularnewline
\hline 
$\tilde{k}$ & $K=t_{0}\tilde{k}\exp(-E)=k/\epsilon$ & Reaction prefactor\tabularnewline
\hline 
$t$ & $\tau=t/t_{0}=\epsilon tu_{a}/x_{0}$ & Time\tabularnewline
\hline 
$X$ & $x=\left(X-D_{0}t\right)/x_{0}-\tau=\left[X-\left(D_{0}+\epsilon u_{0}\right)t\right]/x_{0}$ & Longitudinal direction\tabularnewline
\hline 
$Y$ & $y=\left(\sqrt{2\epsilon}q^{1/4}/x_{0}\right)Y$ & Transverse direction\tabularnewline
\hline 
$\mu,d,\kappa$ & $\nu=2/\left(3\sqrt{q}\epsilon\mathrm{Re}\right)$ & Dissipative effects\tabularnewline
\hline 
\end{tabular}
\par\end{centering}

\caption{\label{tab:Summary-of-parameters}Summary of scaling relationships.}
\end{table}
The relationships between the asymptotic variables $u$, $v$ and
$\lambda$ and the physical (dimensionless, rescaled with the upstream
state) quantities, $\rho,\, U,\, V,\, T$ and $\Lambda$ are given
by
\begin{eqnarray}
\rho & = & 1+\epsilon\sqrt{q}u+O(\epsilon^{3/2}),\label{eq:euler-to-asympt-1}\\
U & = & \epsilon\sqrt{q}u+O(\epsilon^{3/2}),\\
V & = & \epsilon^{3/2}\sqrt{2}q^{3/4}v+O(\epsilon^{2}),\\
T & = & 1+\epsilon^{2}\left(\sqrt{q}u+q\lambda\right)+O(\epsilon^{5/2}),\\
\Lambda & = & \lambda+O(\epsilon).\label{eq:euler-to-asympt-5}
\end{eqnarray}

For the remainder of the paper, we focus exclusively on the inviscid
case such that the asymptotic equations take the form 
\begin{eqnarray}
u_{\tau}+uu_{x}+v_{y} & = & -\frac{1}{2}\lambda_{x},\label{eq:asymptotic-model-inviscid-u}\\
v_{x} & = & u_{y},\label{eq:asymptotic-model-inviscid-v}\\
\lambda_{x} & = & -k(1-\lambda)\exp\left[\theta\left(\sqrt{q}u+q\lambda\right)\right].\label{eq:asymptotic-model-inviscid-lambda}
\end{eqnarray}

We note here that without the chemical reaction, (\ref{eq:asymptotic-model-inviscid-u}-\ref{eq:asymptotic-model-inviscid-lambda})
reduce to 
\begin{eqnarray}
u_{\tau}+\left(\frac{u^{2}}{2}\right)_{x}+v_{y} & = & 0,\label{eq:UTSD-1}\\
v_{x} & = & u_{y},\label{eq:UTSD-2}
\end{eqnarray}
which are canonical equations appearing in the analysis of various
physical phenomena modelled by weakly-nonlinear quasi-planar hyperbolic
waves. In terms of the velocity potential, $\phi$, such that $\nabla\phi=\left(u,v\right)$,
these equations can be rewritten as 
\begin{equation}
\phi_{x\tau}+\left(\frac{\phi_{x}}{2}\right)_{x}^{2}+\phi_{yy}=0,
\end{equation}
which is the well-known unsteady small disturbance transonic equation
\cite{lin1948two}. In nonlinear acoustics, this equation is also
known as the Zabolotskaya-Khokhlov equation \cite{Zab-Khokhlov-1969}. 

In the following sections, we investigate the steady-state solutions
of (\ref{eq:asymptotic-model-inviscid-u}-\ref{eq:asymptotic-model-inviscid-lambda}),
their spectral stability and nonlinear dynamics, and demonstrate that
the asymptotic theory contains the essential features of not only
the steady-state one-dimensional, but also the unsteady and multi-dimensional
travelling waves of the reactive Euler equations.

\section{Travelling wave solutions and their linear stability\label{sec:Traveling-wave-solutions-and-stability}}

In this section, we analyze the travelling wave solutions of the asymptotic
equations and their spectral stability, showing that the asymptotic
solutions are analogous to their ZND counterparts. We start by presenting
the one-dimensional travelling wave solutions in Section \ref{sub:Travelling-wave-solutions},
which are the basis for the one- and multi-dimensional stability analyses
presented in Section \ref{sub:The-spectral-stability}.

\subsection{\label{sub:Travelling-wave-solutions}Travelling wave solutions of
the asymptotic model}

Seeking the one-dimensional travelling wave solutions of the inviscid
asymptotic model (\ref{eq:asymptotic-model-inviscid-u}-\ref{eq:asymptotic-model-inviscid-lambda})
of the form $\bar{u}=\bar{u}(x-\bar{D}\tau)$, we obtain
\begin{equation}
\bar{u}=\bar{D}+\sqrt{\bar{D}^{2}-\bar{\lambda}},
\end{equation}
where $\bar{D}$ is the speed of the wave and the bar denotes the
steady state. It is easily seen that, for the solution to remain real
at the end of the reaction zone, we must choose $\bar{D}\ge1$. The
precise choice of the value is related to the degree of overdrive
of the wave. Even though the overdriven detonations can be included
in the analysis, we focus on the important case of a self sustained
detonation in which the steady state has a sonic point at the end
of the reaction zone. Then, $\bar{D}=1$ and 
\begin{equation}
\bar{u}=1+\sqrt{1-\bar{\lambda}},
\end{equation}
where $\bar{\lambda}$ solves 
\begin{equation}
\bar{\lambda}_{\xi}=-k\left(1-\bar{\lambda}\right)\exp\left(\theta\left(\sqrt{q}\bar{u}+q\bar{\lambda}\right)\right)\label{eq:reaction-rate-function}
\end{equation}
with $\xi=x-\tau$ and boundary condition $\bar{\lambda}(0)=0$ (this
is analogous to the ZND theory, see \cite{fickett2011detonation}).
This solution is used in the analysis that follows. 

\begin{figure}[h]
\centering{}\includegraphics[width=3in,height=2in]{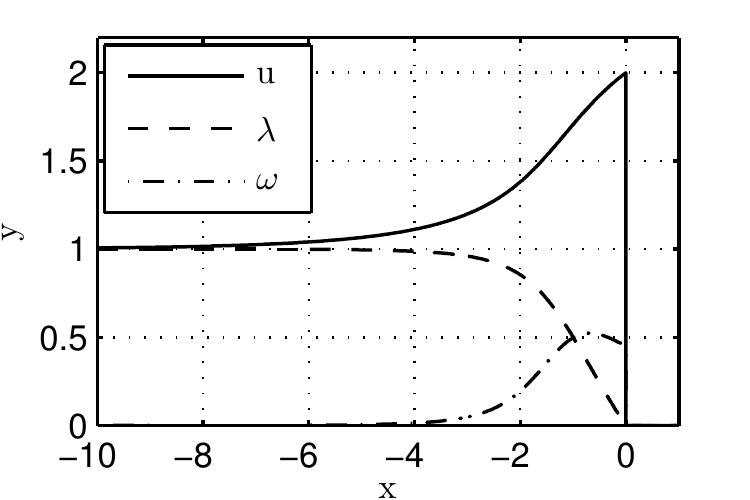}
\caption{\label{fig:Steady-state-profiles}Steady-state profiles and the rate function of the asymptotic solution for $q=1.7$ and $\theta=1.7$.}
\end{figure}

The steady-state structure of the asymptotic model contains substantial
information about the underlying modeling assumptions. A crucial \emph{qualitative}
feature of the structure is that, because of the behaviour of the
leading order correction to temperature, $T_{3}=u+q\lambda$, it is
possible to have a maximum of the reaction-rate function, (\ref{eq:reaction-rate-function}),
inside the reaction zone, as we show in Fig. \ref{fig:Steady-state-profiles}.
The presence of the internal maximum in the reaction rate appears
to be a key factor responsible for the observed unsteady dynamics
of real detonations, consistent with our previous work on a closely
related model equation \cite{kasimov2013model,FariaKasimovRosales-SIAM2014}.
In fact, we have verified that even the qualitative models of Fickett,
\begin{eqnarray}
u_{t}+\left(\frac{u^{2}}{2}+q\lambda\right)_{x} & = & 0,\qquad\lambda_{t}=\omega(\lambda,u),
\end{eqnarray}
and Majda (in the inviscid limit),
\begin{eqnarray}
\left(u+q\lambda\right)_{t}+\left(\frac{u^{2}}{2}\right)_{x} & = & 0,\qquad\lambda_{t}=\omega(\lambda,u),
\end{eqnarray}
are capable of capturing the one-dimensional instabilities of detonation
waves if the reaction rate is taken as (\ref{eq:omega-new}). We recall
that, in these models, the reaction rate usually takes the form 
\begin{equation}
\omega=\left(1-\lambda\right)\varphi\left(u\right),
\end{equation}
 where the ignition function, $\varphi\left(u\right)$, is 
\begin{equation}
\varphi=\begin{cases}
\varphi_{0}\left(u\right), & u>u_{i}\\
0, & u<u_{i}
\end{cases},
\end{equation}
and $u_{i}$ is the ``ignition-temperature'' parameter (see, e.g.,
\cite{Majda:1980zr,humpherys2013stability}). The new rate function,
given by (\ref{eq:omega-new}), reflects the crucially important feature
of the heat-release in unstable detonations, which exhibits a maximum
inside the reaction zone. Thus, both Fickett's and Majda's models
possess the necessary complexity needed to capture the qualitative
dynamics of one-dimensional unstable detonations provided that the
reaction-rate function is chosen appropriately.

\begin{figure}[th]
\centering{}\subfloat[Profiles for $\gamma=1.2,\ Q=0.4,\ E=50$; $\epsilon=1/6$.]{\centering{}\includegraphics[width=3in,height=2in]
{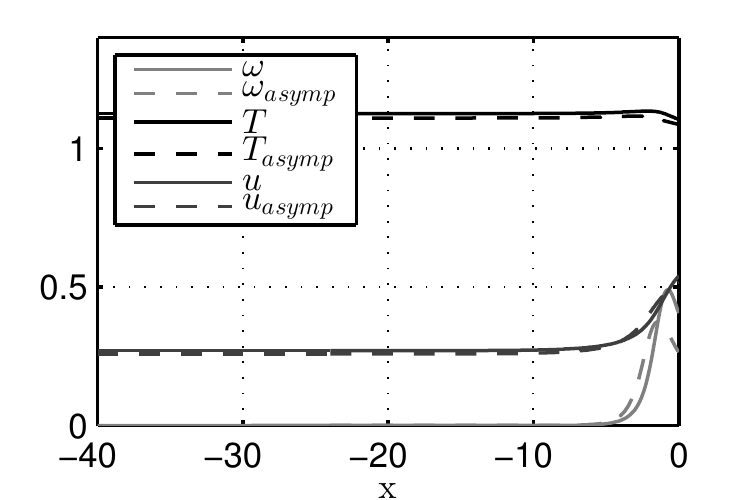}}\subfloat[Profiles for $\gamma=1.1,\ Q=0.4,\ E=50$; $\epsilon=1/11$.]{\centering{}\includegraphics[width=3in,height=2in]{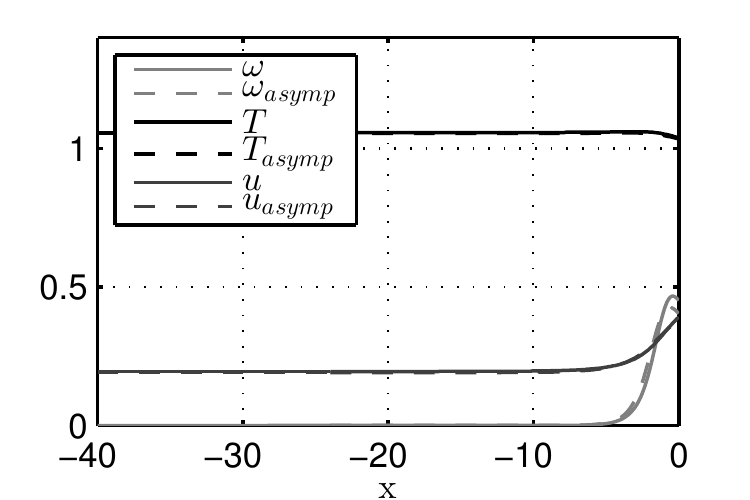}}\\
\subfloat[Error for $\gamma=1.2,\ Q=0.4,\ E=50$; $\epsilon=1/6$.]{\centering{}\includegraphics[width=3in,height=2in]{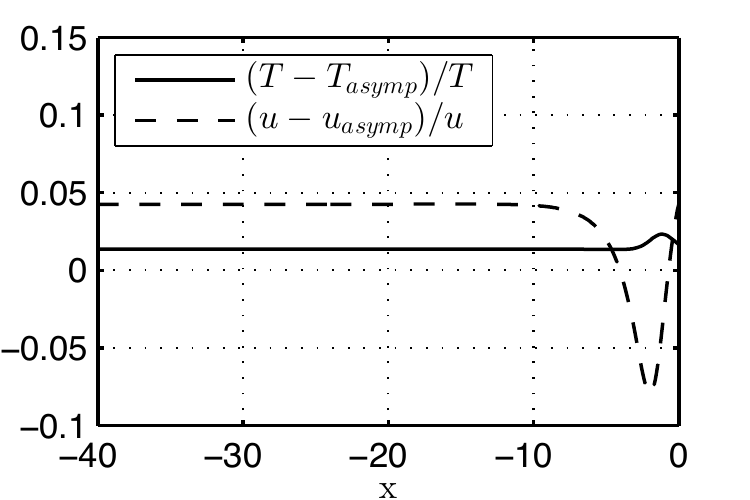}}\subfloat[Error for $\gamma=1.1,\ Q=0.4,\ E=50$; $\epsilon=1/11$.]{\centering{}\includegraphics[width=3in,height=2in]{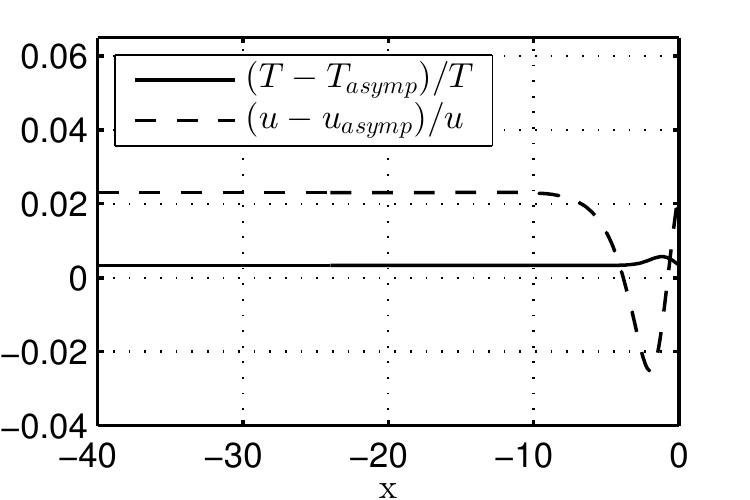}}\caption{\label{fig:Exact-and-asymptotic}Comparison between the exact and
asymptotic steady-state ZND profiles. The asymptotic solutions are
calculated using (\ref{eq:euler-to-asympt-1}-\ref{eq:euler-to-asympt-5}).}
\end{figure}

The steady-state solution also provides a first \emph{quantitative}
test of the accuracy of the asymptotic approximation. In Fig. \ref{fig:Exact-and-asymptotic},
we show a comparison between the ZND solutions of the reactive Euler
equations and their asymptotic counterparts as predicted by the present
theory. We see that the asymptotic approximation performs rather well
when the heat release is small and the activation energy is large,
i.e., when $Q\sim\epsilon$ and $E\sim1/\epsilon^{2}$. For example,
for the realistic value of $\gamma=1.2$ $\left(\epsilon=1/6\right)$,
the relative error is only a small percentage (Fig. \ref{fig:Exact-and-asymptotic}(c)).
As expected, for the smaller value of $\gamma=1.1$ ($\epsilon=1/11$),
the agreement is seen to improve (Fig. \ref{fig:Exact-and-asymptotic}(b,d)),
with the maximum relative error approximately two percent of the flow
velocity. As expected, the approximation worsens as the values of
$\gamma-1$ and $Q$ increase or the value of $E$ decreases.

\subsection{\label{sub:The-spectral-stability}Linear stability theory for the
asymptotic model}

With the steady-state solutions in reasonable agreement with the reactive
Euler equations, we next investigate their stability properties. It
is well known that the reactive Euler equations for the ideal-gas
equation of state and simple-depletion Arrhenius kinetics predict
that the steady-state detonations are unstable when the activation
energy is large enough at a fixed heat release \cite{Erpenbeck64,LeeStewart90,short1998cellular}.
In this section, we analyze the linear stability of the traveling
wave solutions obtained in the previous section to see if the asymptotic
theory agrees with the Euler equations in this regard as well. We
show that, indeed, the steady-state detonation waves become unstable
if either the heat release or the activation energy crosses a certain
threshold. We also demonstrate that multi-dimensional effects play
a nontrivial role in the onset of instabilities.

To proceed with the analysis, let $\bar{u}$ and $\bar{\lambda}$
be the steady-state solution obtained in Section \ref{sub:Travelling-wave-solutions}.
Rewriting (\ref{eq:asymptotic-model-inviscid-u}-\ref{eq:asymptotic-model-inviscid-lambda})
in a shock-attached frame, $\chi=x-s(y,\tau)$, where $s(y,\tau)$
is the shock position, we obtain 
\begin{eqnarray}
u_{\tau}+\left(u-s_{\tau}\right)u_{\chi}+\frac{1}{2}\lambda_{\chi}+v_{y}-s_{y}v_{\chi} & = & 0,\\
u_{y}-s_{y}u_{\chi}-v_{\chi} & = & 0,\\
\lambda_{\chi} & = & -k\left(1-\lambda\right)\exp\left(\theta\left(\sqrt{q}u+q\lambda\right)\right).
\end{eqnarray}
Next, we expand the solution in normal modes,
\begin{eqnarray}
u & = & \bar{u}(\chi)+\delta u_{1}(\chi)\exp\left(\sigma\tau+ily\right)+O\left(\delta^{2}\right),\\
v & = & \delta v_{1}(\chi)\exp(\sigma\tau+ily)+O(\delta^{2}),\\
\lambda & = & \bar{\lambda}(\chi)+\delta\lambda_{1}(\chi)\exp\left(\sigma\tau+ily\right)+O\left(\delta^{2}\right),\\
s & = & \bar{D}\tau+\delta\exp\left(\sigma\tau+ily\right),
\end{eqnarray}
and let $\delta\rightarrow0$. The linearized equations are then 
\begin{eqnarray}
\left(\bar{u}-\bar{D}\right)u_{1}' & = & -(\sigma+\bar{u}')u_{1}+\sigma\bar{u}'-ilv_{1}-\frac{1}{2}\left(g(\chi)u_{1}+h(\chi)\lambda_{1}\right),\label{eq:linear-stability-u}\\
v_{1}' & = & ilu_{1}-il\bar{u}',\label{eq:linear-stability-v}\\
\lambda_{1}' & = & g(\chi)u_{1}+h(\chi)\lambda_{1},\label{eq:linear-stability-lam}
\end{eqnarray}
where the prime denotes a differentiation with respect to $\chi$
and 
\begin{eqnarray}
g(\chi) & = & -\frac{\partial\omega}{\partial u}\left(\bar{u},\bar{\lambda}\right)=-k\theta\sqrt{q}\left(1-\bar{\lambda}\right)\exp\left[\theta\left(\sqrt{q}\bar{u}+q\bar{\lambda}\right)\right],\\
h(\chi) & = & -\frac{\partial\omega}{\partial\lambda}\left(\bar{u},\bar{\lambda}\right)=-k\left[\theta q\left(1-\bar{\lambda}\right)-1\right]\exp\left[\theta\left(\sqrt{q}\bar{u}+q\bar{\lambda}\right)\right].
\end{eqnarray}

The boundary conditions for (\ref{eq:linear-stability-u}-\ref{eq:linear-stability-lam})
are obtained from linearizing (\ref{eq:asymptotic-jump-condition1}-\ref{eq:asymptotic-jump-condition3}):
\begin{equation}
u_{1}(0)=2\sigma;\qquad v_{1}(0)=-2il;\qquad\lambda_{1}(0)=0.
\end{equation}
Noticing that for self sustained detonation, $\bar{u}-\bar{D}\to0$
as $\chi\rightarrow-\infty$, we require that the right-hand side
of (\ref{eq:linear-stability-u}) vanish in the limit as well, i.e.,
\[
H(\sigma,l)=-(\sigma+\bar{u}')u_{1}+\sigma\bar{u}'-ilv_{1}-\frac{1}{2}\left(g(\chi)u_{1}+h(\chi)\lambda_{1}\right)\rightarrow0\quad\text{as }\chi\rightarrow-\infty.
\]
Because $\bar{u'}\rightarrow0,\ g(\chi)\rightarrow0$ as $\chi\rightarrow-\infty$,
this solvability condition (alternatively called the ``boundedness''
or the ``radiation'' condition \cite{LeeStewart90}) simplifies
to 
\begin{equation}
-\sigma u_{1}-ilv_{1}-\frac{1}{2}h\left(-\infty\right)\lambda_{1}\to0\quad\text{as }\chi\to-\infty.\label{eq:radiation-condition}
\end{equation}
To eliminate the numerical inconvenience of $\sigma=0,\ l=0$ always
being an eigenvalue -- a consequence of the translation invariance
of the traveling wave -- we rescale $u_{1}$ and $\lambda_{1}$ by
$\sigma$ and $v_{1}$ by $il$. The stability problem is then posed
as follows: 

Solve
\begin{eqnarray}
\left(\bar{u}-\bar{D}\right)u_{1}' & = & -(\sigma+\bar{u}')u_{1}+\bar{u}'+\frac{l^{2}}{\sigma}v_{1}-\frac{1}{2}\left(g(\chi)u_{1}+h(\chi)\lambda_{1}\right),\label{eq:spectral-stability-eq1}\\
v_{1}' & = & \sigma u_{1}-\bar{u}',\label{eq:spectral-stability-eq2}\\
\lambda_{1}' & = & g(\chi)u_{1}+h(\chi)\lambda_{1},\label{eq:spectral-stability-eq3}
\end{eqnarray}
subject to $u_{1}(0)=2,\ v_{1}(0)=-1$ and $\lambda_{1}(0)=0$ at
the shock and the boundedness condition (\ref{eq:radiation-condition})
at negative infinity.

The preceding eigenvalue problem is solved numerically using the shooting
method of \cite{LeeStewart90}. We solve the problem for different
values of $\theta$, $q$ and $l$, which are the only remaining parameters.
Consistent with the behaviour of the stability spectrum of detonation
waves in reactive Euler equations \cite{LeeStewart90}, we find that
unstable modes do exist either for large enough $q$ or for large
enough $\theta$. We also find that the transverse modes, where $l\neq0$,
tend to be more unstable than purely longitudinal disturbances \cite{short1998cellular}.

\begin{figure}[h]
\begin{centering}
\subfloat[$q=1.8$]{\includegraphics[width=2in,height=2in]{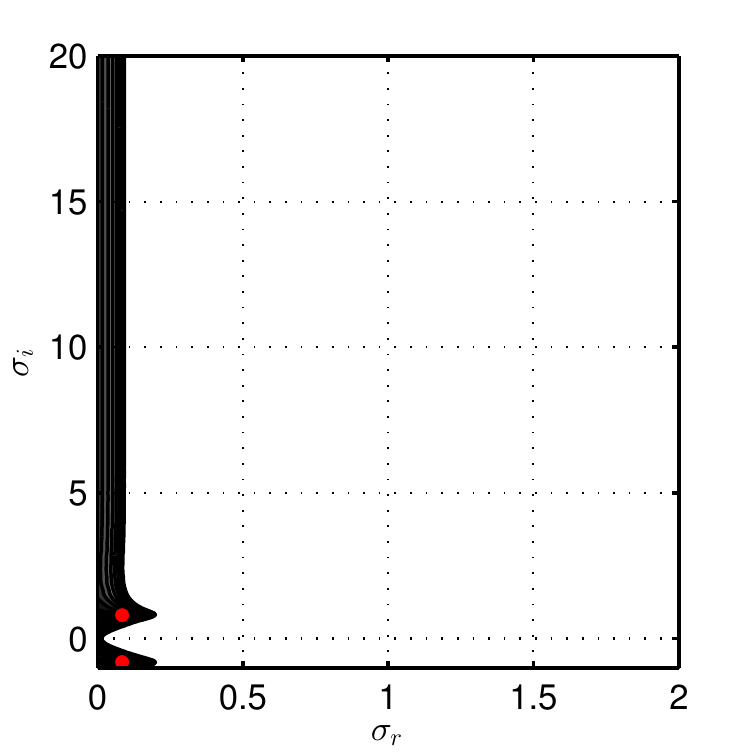}}\subfloat[$q=2.2$]{\includegraphics[width=2in,height=2in]{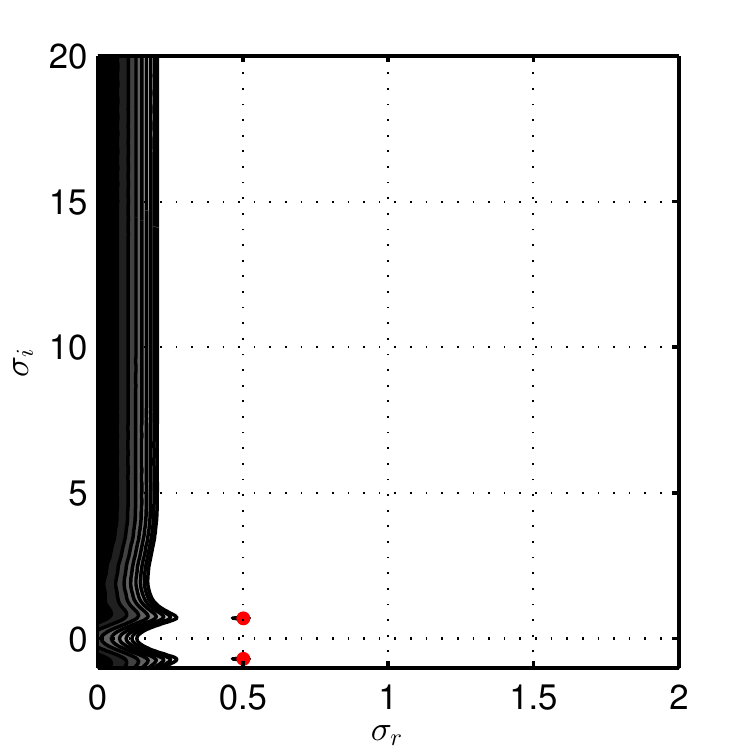}}\subfloat[$q=2.6$]
{\includegraphics[width=2in,height=2in]{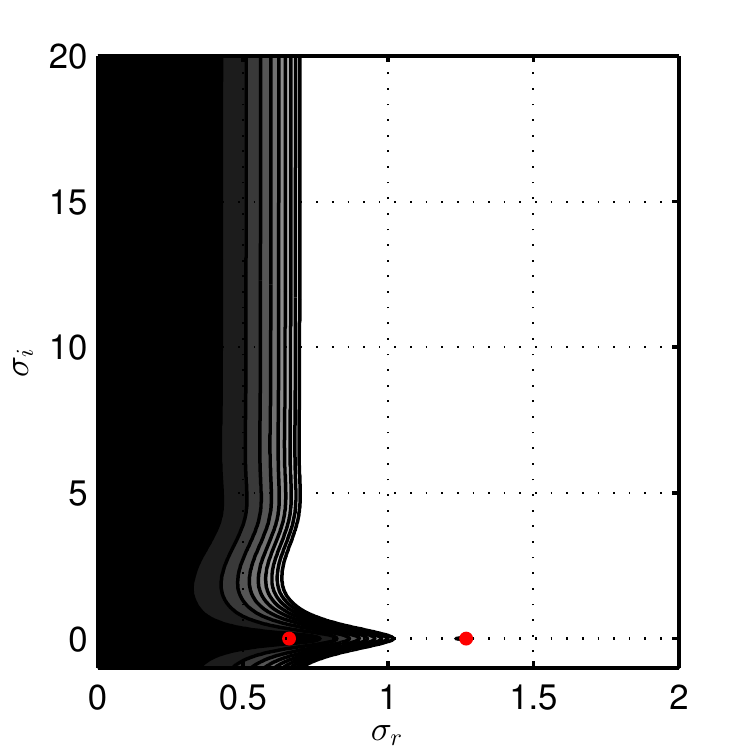}}\\
\subfloat[$q=2.75$]{\includegraphics[width=2in,height=2in]{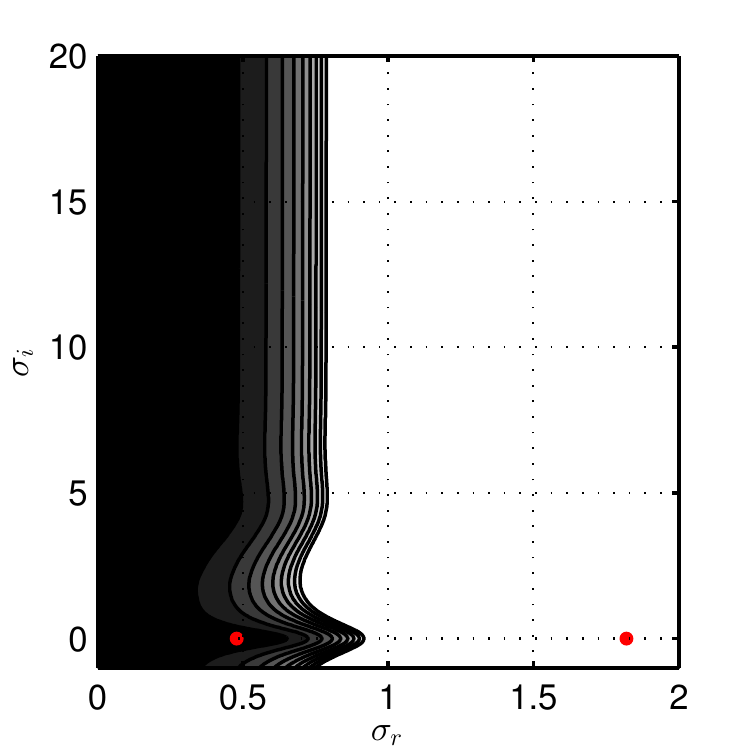}}\subfloat[$q=3.5$]{\includegraphics[width=2in,height=2in]{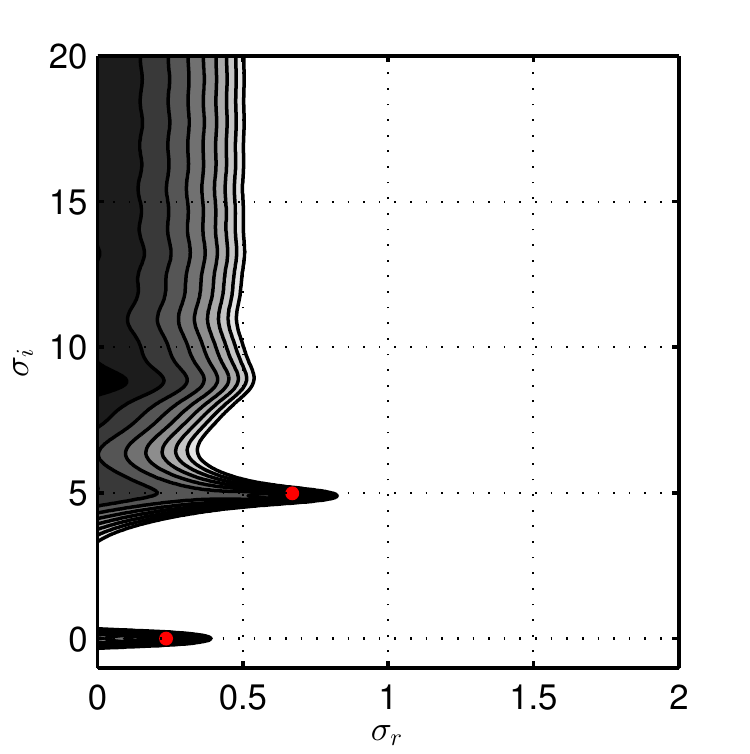}}\subfloat[$q=4.5$]
{\includegraphics[width=2in,height=2in]{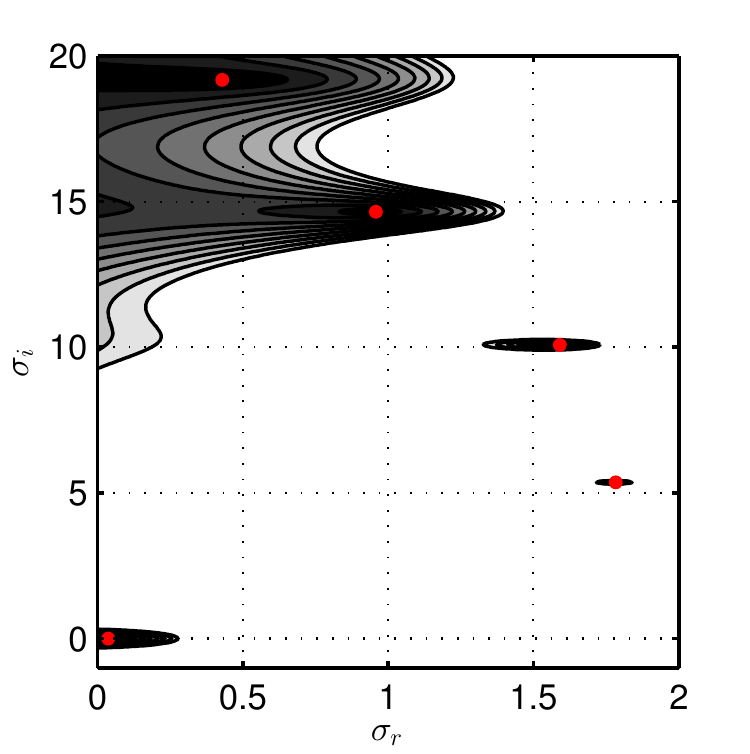}}
\par\end{centering}

\caption{\label{fig:Spectrum}Contour plots of the spectral function (\ref{eq:H-1D})
for $\theta=1.8$ and increasing $q$. Note that the real part of
the non-oscillatory root increases with $q$. The red dots in the
figures represent the eigenvalues. In (e) and (c), the dominant non-oscillatory
root is not shown. }
\end{figure}

First, we consider the purely one-dimensional problem, i.e., with
$l=0$. In Fig. \ref{fig:Spectrum}, we show the contour plot of the
absolute value of the stability function, 
\begin{equation}
|H\left(\sigma,0\right)|=\left|-(\sigma+\bar{u}')u_{1}+\sigma\bar{u}'-\frac{1}{2}\left(g(\chi)u_{1}+h(\chi)\lambda_{1}\right)\right|,\label{eq:H-1D}
\end{equation}
as a function of real and imaginary parts of $\sigma$. The valleys
in the plot of $|H(\sigma,0)|$, which correspond to the darker regions
in Fig. \ref{fig:Spectrum}, provide an initial guess for the location
of eigenvalues. A root solver is then used wherein the complex function,
$H(\sigma,0)$, is set to zero in order to accurately locate the eigenvalues.\textbf{
}An increasing number of unstable eigenvalues is seen as the neutral
boundary is crossed by increasing the heat release, $q$, for a given
value of the activation energy, $\theta$. The qualitative behavior
of the spectrum is in agreement with that known for the Euler detonations
\cite{LeeStewart90}. Furthermore, Figs. \ref{fig:Spectrum}(a-d)
show the migration of the oscillatory complex conjugates, with $\sigma_{i}\neq0$,
into non-oscillatory unstable modes that are also observed in the
Euler equations \cite{short1997multidimensional}. Finally, in Figs.
\ref{fig:Spectrum}(e,f), we see that far into the unstable regime,
many eigenvalues are found, indicating a complexity of the linear
spectrum.

\begin{figure}[h]
\centering{}\subfloat[\label{fig:Neutral-stability-boundary}Neutral stability boundaries
for the first two low-frequency modes in the asymptotic model. The
modes are unstable to the right of the curves. ]{\noindent \begin{centering}
\includegraphics[width=3in,height=2in]{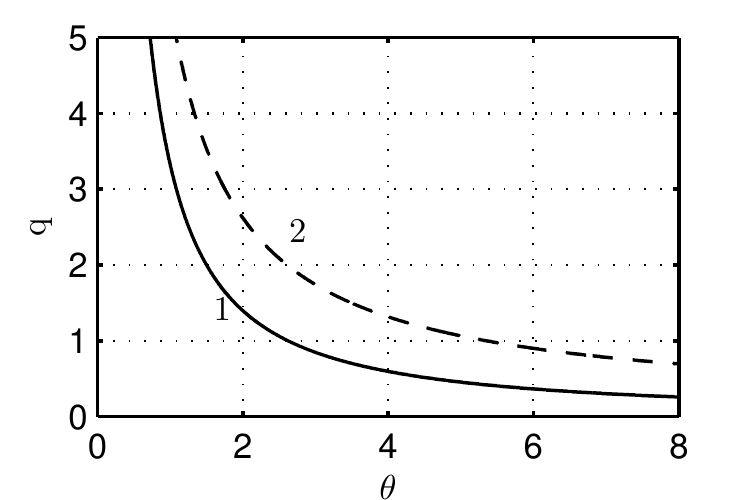}
\par\end{centering}

}\hfill\subfloat[\label{fig:Comparisons-with-L&S}Comparison of the asymptotic and
the exact \cite{LeeStewart90} neutral stability boundaries for $\gamma=1.2$,
in the heat release, $Q$, versus activation energy, $E$, plane. ]{\begin{centering}
\includegraphics[width=3in,height=2in]{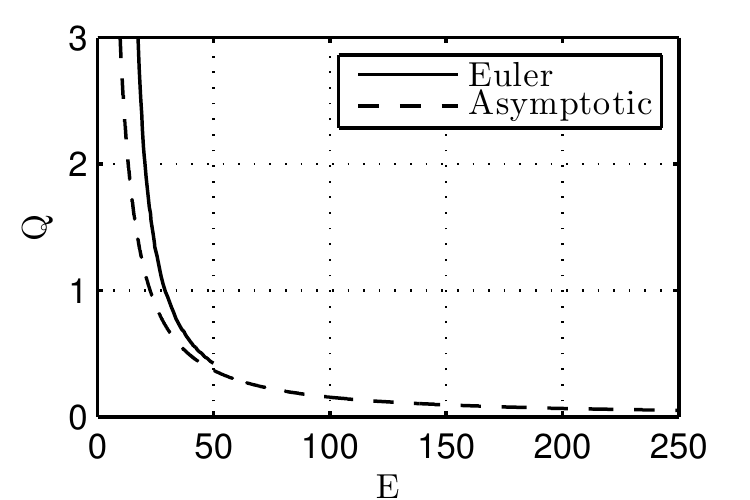}
\par\end{centering}

}\caption{\label{fig:Neutral-stability-curves}Neutral stability curves.}
\end{figure}

We also obtain the neutral stability curves for the first two unstable
eigenvalues in the asymptotic model, as shown in Fig. \ref{fig:Neutral-stability-curves}(a).
It is seen that the lowest frequency mode $1$ is more unstable than
mode $2$ for a wide range of $q$ and $\theta$. Substantially away
from the neutral boundary, the non-oscillatory root may become dominant,
as is seen in the example shown in Fig. \ref{fig:Spectrum}(c,d).

In Fig. \ref{fig:Neutral-stability-curves}(b), we plot the neutral
curve for the lowest frequency eigenvalue, indicated by the solid
line in Fig.\textbf{ }\ref{fig:Neutral-stability-curves}(a), in the
plane of the heat release, $Q$, and the activation energy, $E$.
The result is compared with the neutral curve computed directly from
the reactive Euler equations \cite{LeeStewart90} and a reasonably
close agreement between the two is seen. We observe that, as expected,
the agreement improves with smaller $Q$ and larger $E$.\textbf{
}Finally, since the asymptotic model allows for easy calculations
of the high activation energy/small heat release limit, we extend
the prediction of the neutral boundary to rather high values of $E\approx250$\textbf{
}and note that the asymptotic curve follows the scaling $Q\sim1/E$
very well. In fact, a simple least-squares fitting of the form $Q=c_{1}/E$
gives an $r^{2}$ value of $0.9841$.

\begin{figure}[h]
\centering{}%
\begin{tabular}{cc}
\includegraphics[width=3in,height=2in]{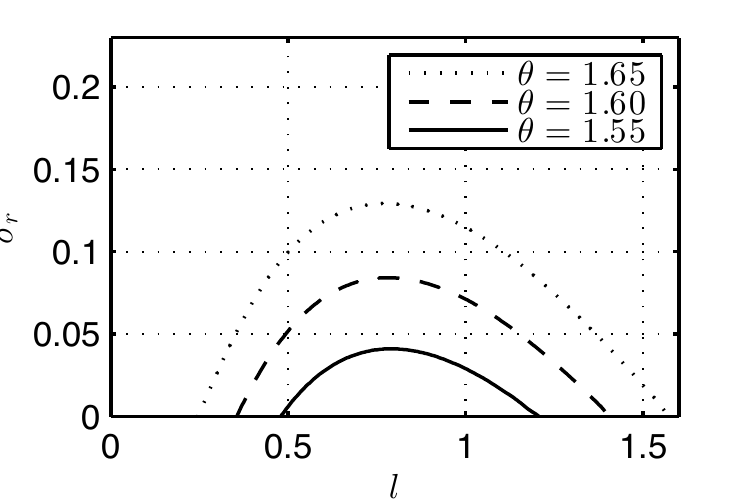} & \includegraphics[width=3in,height=2in]{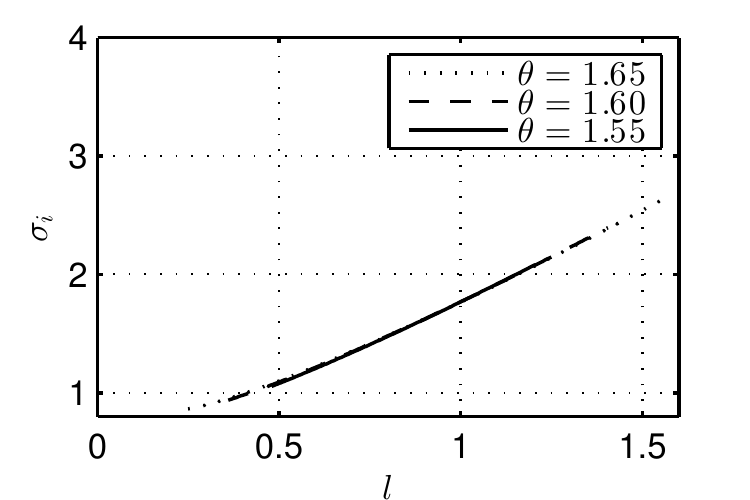}\tabularnewline
\end{tabular}\caption{\label{fig:Spectrum-2d}Growth rate and frequency for the most unstable
mode versus the wave-number, $l$, for several values of the activation
energy, $\theta$. }
\end{figure}

If $l\neq0$, then there is the possibility that transverse waves
will trigger the instabilities. This occurs in the Euler equations,
where it is found that multi-dimensional instability prevails over
purely longitudinal instability. Again, using the shooting method,
we solve (\ref{eq:spectral-stability-eq1}-\ref{eq:spectral-stability-eq3})
numerically for various values of $l$. Solving for the roots of the
radiation function, 
\begin{equation}
H(\sigma,l)=-(\sigma+\bar{u}')u_{1}+\sigma\bar{u}'-ilv_{1}-\frac{1}{2}\left(g(\chi)u_{1}+h(\chi)\lambda_{1}\right),
\end{equation}
at $\chi=-\infty$, we obtain the two-dimensional stability spectrum.
We first fix $q=1.7$ and vary $\theta=1.65,\,1.60,\,1.55$. In Fig.
\ref{fig:Spectrum-2d}, we show the real and imaginary parts of the
unstable modes for relatively small values of $l$. It is seen that
the model predicts purely two-dimensional instabilities for a certain
range of the transverse wave numbers, and that increasing the activation-energy
parameter, $\theta$\textbf{,} has a destabilizing effect on the steady-state
solutions. 

\begin{figure}[h]
\centering{}%
\begin{tabular}{cc}
\includegraphics[width=3in,height=2in]{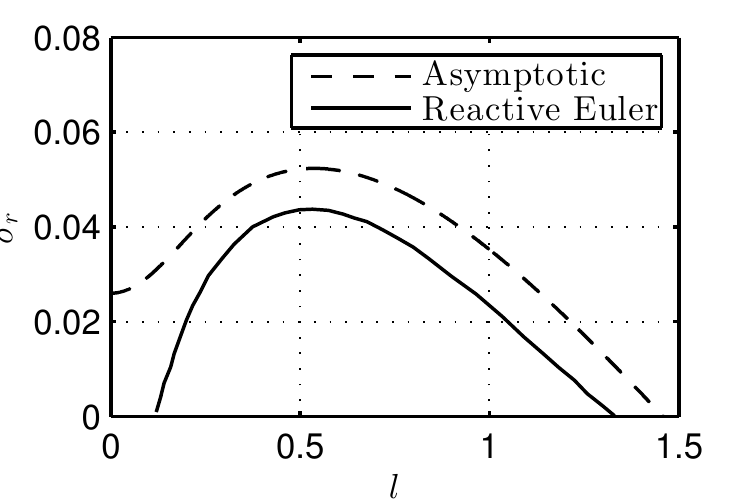} & \includegraphics[width=3in,height=2in]{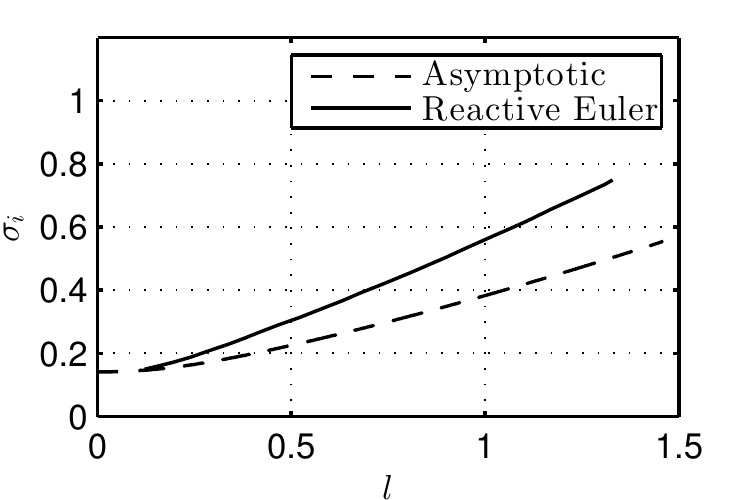}\tabularnewline
\end{tabular}\caption{\label{fig:Spectrum-2d-euler-vs-asymptotic}Comparisons of the growth
rate and frequency for the most unstable mode versus the wave number,
for $Q=0.4$ and $E=50$. The dashed curve corresponds to the model
in this paper. The solid line corresponds to the reactive Euler equations
as computed in \cite{taylor2009mode}.}
\end{figure}

We also perform a quantitative comparison between the two-dimensional
stability of the asymptotic model and the known stability diagram
for the Euler equations. We choose the values of $\gamma=1.2,\ Q=0.4$
and $E=50$. Then, after performing the appropriate conversion between
dimensionless variables, we compare the asymptotic results with those
obtained in \cite{taylor2009mode} (see Fig. \ref{fig:Spectrum-2d-euler-vs-asymptotic}).
We observe a fair agreement. There are, however, some differences.
We see that for the parameters chosen in Fig. \ref{fig:Spectrum-2d-euler-vs-asymptotic},
the asymptotic model predicts one-dimensional instabilities, while
the reactive Euler equations do not. Also, we observe that the disagreement
between the imaginary parts of the eigenvalues increases with increasing
wavenumber. This should be expected because short transverse wavelengths
cannot be represented accurately in the weak curvature limit assumed
for this model. In the next section, we investigate the long-time
non-linear dynamics of the asymptotic solutions in regimes corresponding
to linearly unstable steady-state one-dimensional solutions. The calculations
are performed in both one and two spatial dimensions.

\section{Nonlinear dynamics of the asymptotic model\label{sec:Nonlinear-dynamics}}

In the previous section, we showed that the asymptotic model exhibits
the same linear stability behavior as the reactive Euler equations.
The question of what happens after the onset of instabilities can
be investigated through direct numerical simulations of the model
equations. We show below that the traveling wave solutions of the
asymptotic model reproduce, in both one and two spatial dimensions,
the complexity observed in solutions of the reactive Euler equations.

\subsection{Galloping detonations \label{sub:Galloping-detonations}}

We focus here on the ability of the asymptotic model to predict the
complex nonlinear dynamics of pulsating (galloping) detonations. We
perform a detailed numerical investigation of the large time asymptotic
behaviour of oscillatory solutions of the model. In the one-dimensional
inviscid case, the system given by (\ref{eq:asymptotic-model-u}-\ref{eq:asymptotic-model-lambda})
reduces to 
\begin{eqnarray}
u_{\tau}+uu_{x} & = & -\frac{1}{2}\lambda_{x},\label{eq:1d-asymptotic-1}\\
\lambda_{x} & = & -k(1-\lambda)\exp(\theta(\sqrt{q}u+q\lambda)).\label{eq:1d-asymptotic-2}
\end{eqnarray}
This system resembles the one derived in \cite{RosalesMajda:1983ly}
with one crucial difference -- as a consequence of the $\gamma-1=O(\epsilon)$
assumption, the reaction rate function in (\ref{eq:1d-asymptotic-2})
has a more complicated $\lambda$-dependence, which is in fact at
the heart of the complexity of the solutions obtained here. System
(\ref{eq:1d-asymptotic-1}-\ref{eq:1d-asymptotic-2}) is also the
same as in \cite{clavin2002dynamics}, where the $\gamma-1=O(\epsilon)$
assumption is used.

\begin{figure}[h]
\begin{centering}
\subfloat[Stability of the ZND solution at $\theta=1.705<\theta_{c},\ q=1.7$.]{\centering{}\includegraphics[width=6in,height=2in]{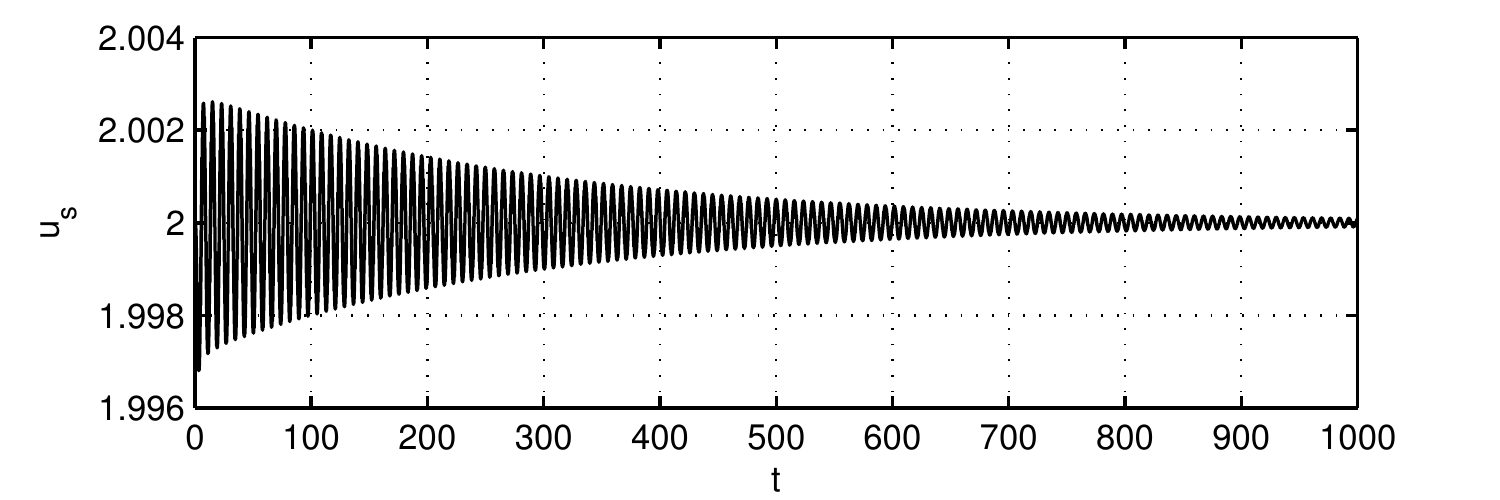}}
\par\end{centering}

\begin{centering}
\subfloat[Instability of the ZND solution and the limit-cycle attractor at $\theta=1.715>\theta_{c},\ q=1.7$.
Notice that the amplitude of the limit cycle scales roughly as the
distance to the bifurcation point, as in a super-critical Hopf bifurcation. ]{\centering{}\includegraphics[width=6in,height=2in]{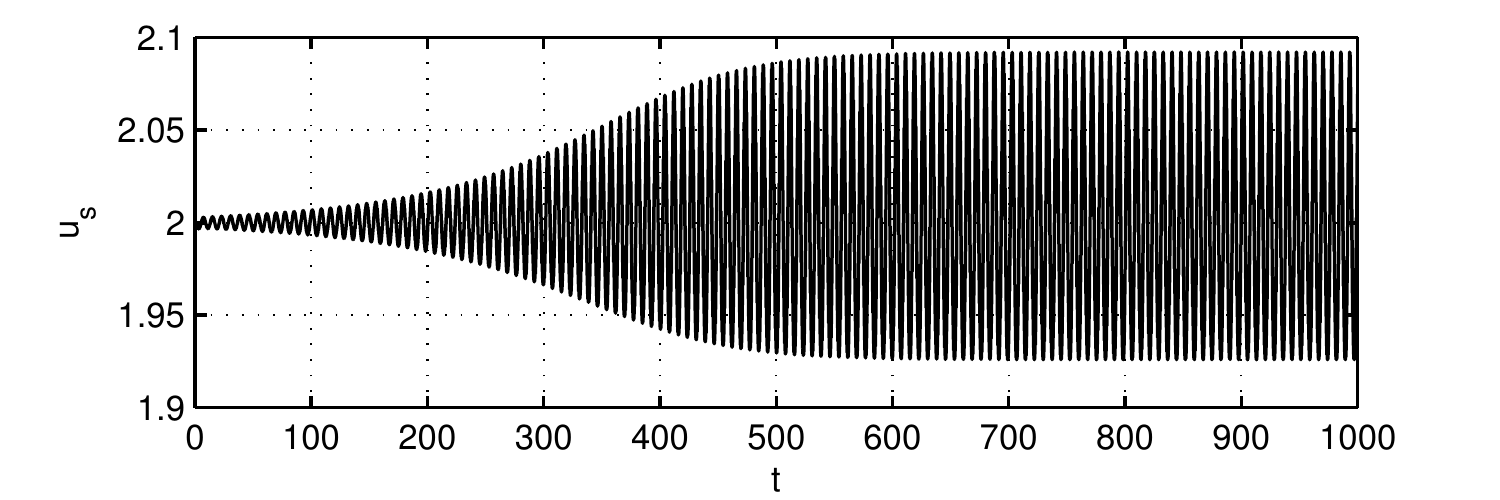}}
\par\end{centering}

\centering{}\caption{\label{fig:Nonlinear-dynamics}Nonlinear dynamics of (\ref{eq:1d-asymptotic-1}-\ref{eq:1d-asymptotic-2})
near the neutral boundary, $\theta_{c}=1.710$, as predicted by the
linear stability theory for $q=1.7$.}
\end{figure}

We solve (\ref{eq:1d-asymptotic-1}-\ref{eq:1d-asymptotic-2}) numerically
in a shock-attached frame \cite{kasimov2004dynamics}, using a second-order
finite volume scheme with a second-order Total Variation Diminishing
(TVD) Runge-Kutta temporal discretization \cite{leveque2002finite}.
Because no differentiation across the shock is performed, true second-order
convergence is obtained for the cases tested, which include the convergence
to various stable steady-state ZND solutions. We further verified
the numerical algorithm by performing a cross-validation between the
linear stability solver and the direct numerical solver. That is,
we made sure that for several values of $q$, the linear stability
prediction of the neutral boundary agrees with the neutral stability
boundary of the numerical scheme for the nonlinear model. For example,
when $q=1.7$, the linear stability curve in Fig. \ref{fig:Neutral-stability-curves}(a)
indicates that $\theta_{c}=1.710$ is the neutral value of the activation
energy, i.e., the ZND wave is unstable for $\theta>\theta_{c}$ and
stable for $\theta<\theta_{c}$. Numerical simulations of (\ref{eq:1d-asymptotic-1}-\ref{eq:1d-asymptotic-2})
with $\theta$ slightly above and slightly below $\theta_{c}$ confirm
this prediction, as shown in Fig. \ref{fig:Nonlinear-dynamics}, where
the shock state, $u_{s}$, is plotted as a function of time. 

\begin{figure}[h]
\centering{}\subfloat[$\theta=0.75$]{\includegraphics[width=3in,height=2in]{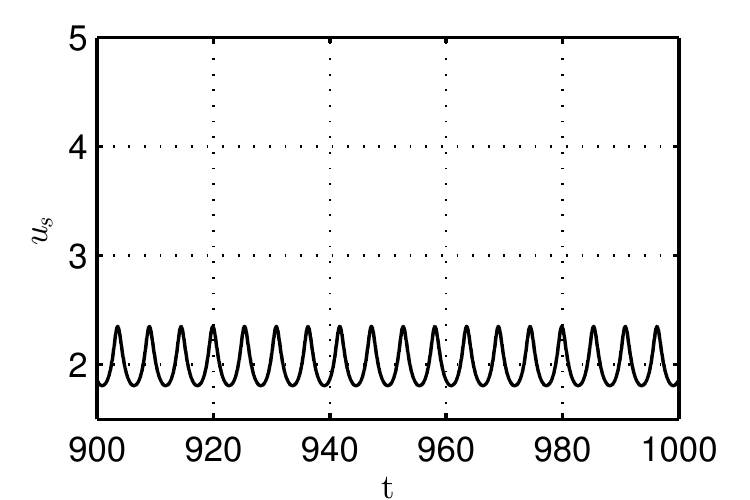}}\hfill\subfloat[$\theta=0.8$]{\includegraphics[width=3in,height=2in]{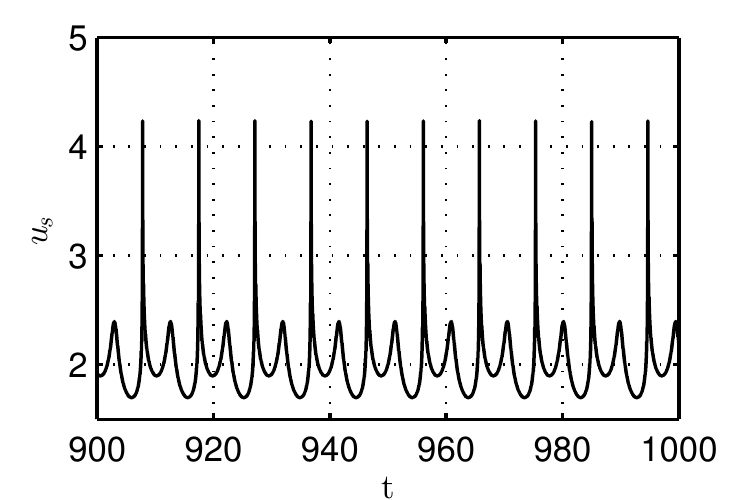}}\\
\subfloat[$\theta=0.83$]{\includegraphics[width=3in,height=2in]{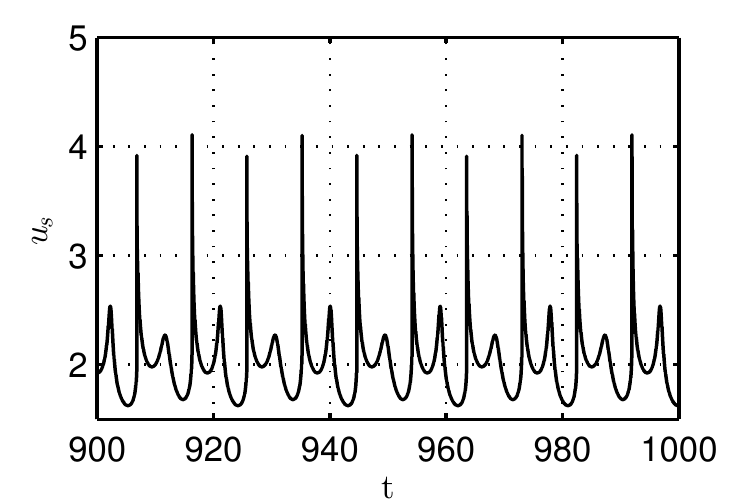}}\hfill\subfloat[$\theta=0.85$]{\includegraphics[width=3in,height=2in]{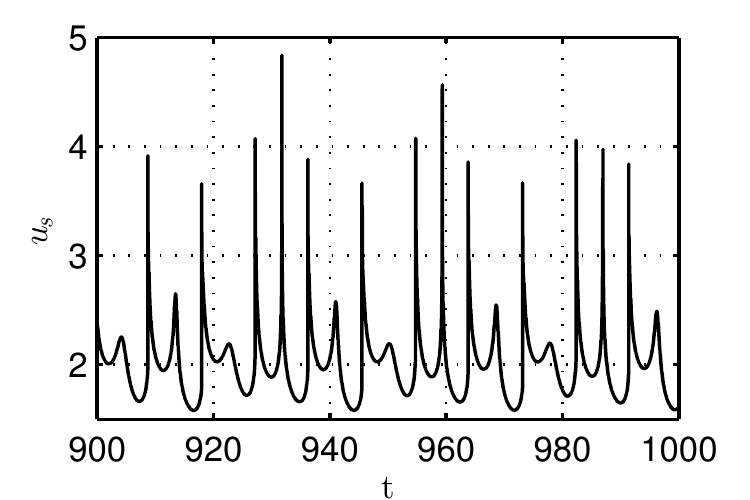}}\caption{\label{fig:The-shock-state}The shock state as a function of time
for increasing values of $\theta$ and fixed $q=5$ showing pulsations
of different complexity. }
\end{figure}

We also compute solutions of (\ref{eq:1d-asymptotic-1}-\ref{eq:1d-asymptotic-2})
further away from the neutral boundary in order to check if the model
captures a sequence of bifurcations leading to chaos as occurs in
the reactive Euler equations \cite{Ng2005,HenrickAslamPowers2006}.
Such a sequence of bifurcations is indeed present in the model. Long
time simulations show that the solutions tend to either a fixed point,
a limit cycle or (what appears to be) a chaotic attractor. We run
the simulations at $q=5$ and plot the post-shock state, $u_{s}$
(by the Rankine-Hugoniot conditions, $u_{s}=2D$, where $D$ is the
shock speed), as a function of time for several different types of
solution, as shown in Fig. \ref{fig:The-shock-state}. Beyond the
stability boundary, the shock velocity becomes oscillatory. Near the
neutral boundary, the oscillations have a small amplitude and are
periodic (Fig. \ref{fig:The-shock-state}(a)), but the structure of
each period becomes more complex as we move away from the neutral
boundary by increasing the activation energy (Fig. \ref{fig:The-shock-state}(b,c)).
Eventually, a value of $\theta$ is reached at which no obvious period
is evident as seen in Fig. \ref{fig:The-shock-state}(d). 

\begin{figure}[H]
\centering{}\includegraphics[width=6in,height=4in]{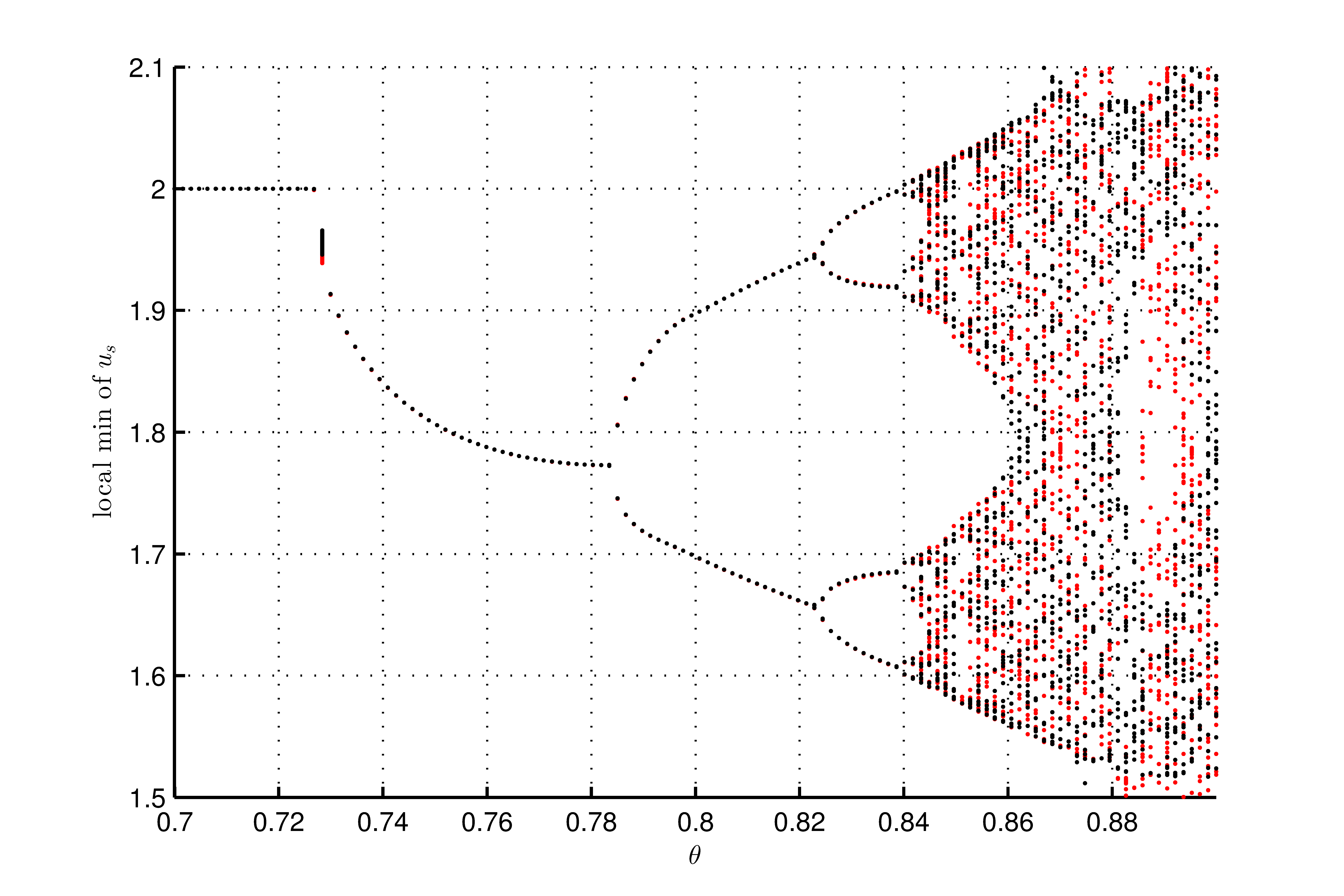}\caption{\label{fig:convergence-bif-diagram}
The bifurcation diagram at $q=5$ showing the local minima of the attractor solution's shock value,
$u_{s}(\tau)$, as a function of the activation energy, $\theta$. The simulations were carried out at
$N=15,000$ (red) and $N=30,000$ (black) grid points on the computational domain of length $L=30$. Away from the chaotic regimes, the predictions are seen to be nearly identical.}
\end{figure}

The behaviour described in the previous paragraph can be understood
in terms of a period doubling sequence of bifurcations where the period
becomes longer and more complex with each bifurcation. In order to
construct a bifurcation diagram illustrating this process, we proceed
as follows: for each value of $\theta$, we follow the evolution of
the solution until it settles on the attractor. Then, we extract the
set of local minima for $u_{s}(\tau)$ -- a finite set for any periodic
solution. The values of these minima are then plotted versus the bifurcation
parameter, $\theta$. The result is shown in Fig. \ref{fig:convergence-bif-diagram},
which is reminiscent of the standard Feigenbaum period doubling cascade
leading to chaos \cite{strogatz1994nonlinear}.

It is important to note that the further we move into the unstable
region, the harder it is to numerically capture the wave dynamics
with good accuracy. That is, in the highly unstable regime ($\theta\gtrapprox0.85$
in Fig. \ref{fig:convergence-bif-diagram}), the quantitative details
of the bifurcation diagram are sensitively dependent on the grid resolution.
In a truly chaotic regime, such sensitivity is intrinsic and reflects
the nature of the system. However, another reason, which is at play
even before the apparently chaotic regime sets on, is that the wave
dynamics can involve spatial scales that undergo large changes (by
orders of magnitude) during the wave evolution. This is a direct consequence
of the Arrhenius exponential dependence of the reaction rate, which
can trigger large variations in the reaction rate from moderate changes
in the temperature when the activation energy is large. This issue
of stiffness associated with high-activation energy detonations in
unstable regimes is discussed next in more detail.

In theoretical and numerical studies of detonation, a widely used
spatial scale is taken to be the half-reaction length, $x_{1/2}$,
defined as the distance from the shock where half of the energy is
released in the ZND solution. The space is then non-dimensionalized,
as done in this work, so that the half-reaction happens over a unit
length. The numerical resolution is thus measured as a number of points
per this unit of length. Although appropriate for stable or weakly
unstable detonations, the ZND half-reaction length and therefore the
resolution measured on this scale become less meaningful when considering
unstable detonations at high activation energies. The reason is that
the energy release can become extremely localized during the dynamical
evolution of pulsating waves and therefore the actual number of grid
points used to capture the heat release region can significantly decrease.
\begin{figure}[H]
\centering{}\subfloat[$\theta=0.75$]{\includegraphics[width=3in,height=2in]{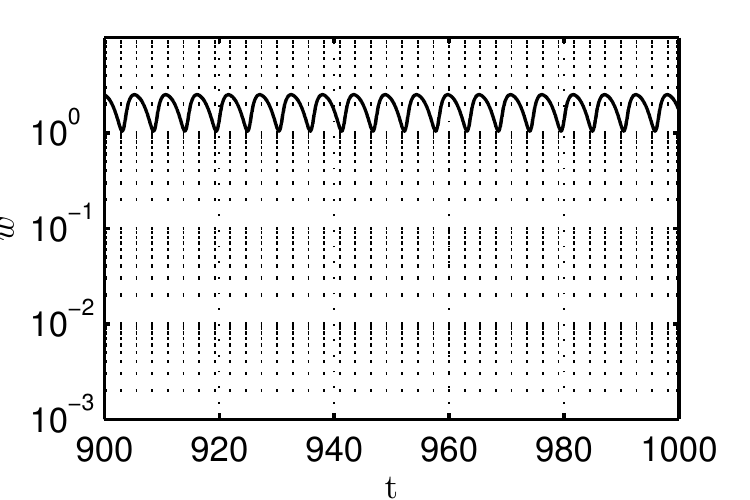}}\hfill\subfloat[$\theta=0.8$]{\includegraphics[width=3in,height=2in]{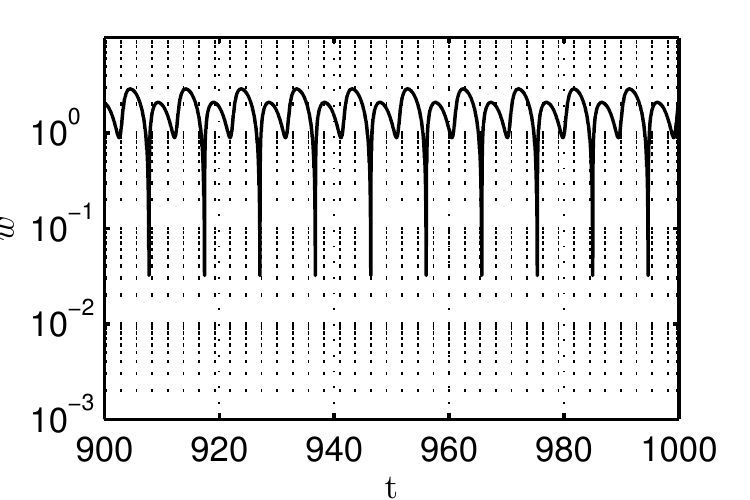}}\\
\subfloat[$\theta=0.83$]{\includegraphics[width=3in,height=2in]{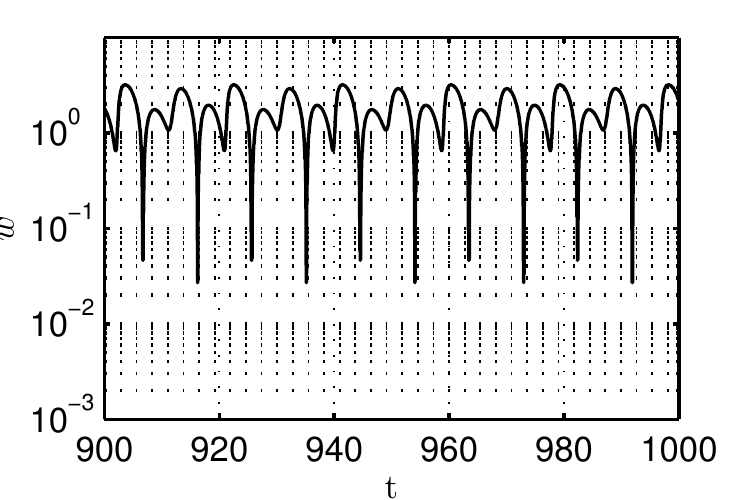}}\hfill\subfloat[$\theta=0.85$]{\includegraphics[width=3in,height=2in]{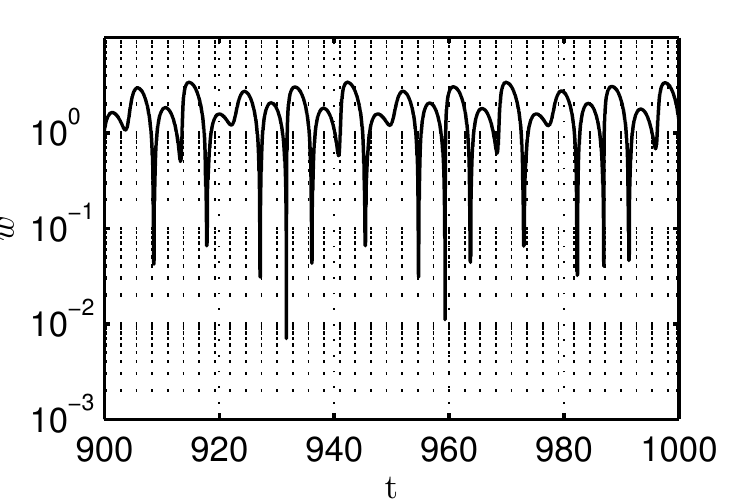}}\caption{\label{fig:width-of-reaction-zone}The width of the reaction zone
as a function of time. The simulations were performed in a domain
of length $L=30$ and with $N=30,000$ grid points ($dx=10^{-3}$).}
\end{figure}

In order to provide a more quantitative measure of the variations
of relevant spatial scales during the time evolution of a pulsating
detonation, we introduce the width of the reaction zone as the smallest
value, $w_{m}$, such that there exists an interval, $\mathcal{I}$,
of length $w_{m}$, where $m$ percent of the energy is released.
For nice enough rate functions, $w_{1/2}$ of the ZND solution is
roughly equivalent to the half-reaction length, $x_{1/2}$. Notice,
however, that for chemical reactions with large induction zones and/or
localized heat release, the values of $w_{1/2}$ and $x_{1/2}$ are
significantly different, with $w_{1/2}$ becoming a more relevant
spatial scale. In the calculations below, we use $m=0.95$ such that
$w_{0.95}$ represents, at a given time, the smallest width containing
$95\%$ of the heat release.

In Fig. \ref{fig:width-of-reaction-zone}, we show $w_{0.95}$ as
a function of time during the detonation evolution. We fix the numerical
resolution at $1000$ points per half-reaction length ($dx=0.001$),
which may be considered an overkill for a steady ZND wave. We then
compute $w_{0.95}$ as a function of time at different activation
energies. As can be seen in Fig. \ref{fig:width-of-reaction-zone}(a),
in the weakly unstable regime, the width of the heat-release zone
changes by a factor of about two during the time evolution of the
wave; the heat release region is still well resolved. Further into
the unstable regime (Fig. \ref{fig:width-of-reaction-zone}(b,c)),
we see that the relevant size of the reaction zone can shrink by more
than an order of magnitude and, thus, even with $1000$ points per
half-reaction length, there are times when only about $30$ points
are used to resolve the heat-release region. If we increase the activation
energy even further, stepping into the apparently chaotic regime,
we observe very short windows of time when only about $10$ points
are being used per heat-release zone. 

Because of the difficulties outlined above, without resorting to adaptive
mesh refinement, $500-1000$ points per half-reaction length are needed
to obtain a bifurcation diagram with features that are essentially
grid independent away from the chaotic regimes (see Fig. \ref{fig:convergence-bif-diagram}).
While recognizing that such sensitivity to initial conditions or discretization
errors is natural for chaotic dynamics, caution is still required
when interpreting the results of numerical simulations of unstable
detonations with high activation energies.

\subsection{Cellular detonation\label{sub:Cellular-detonation}}

Multi-dimensional instability is very important in gaseous detonations
and results in cellular structures involving triple-point interactions
on the detonation lead shock \cite{OranBoris,fickett2011detonation,Lee-2008}.
It is therefore crucial to check if the asymptotic model, (\ref{eq:asymptotic-model-u}-\ref{eq:asymptotic-model-lambda}),
can reproduce the dynamics of not only one-dimensional, but also multi-dimensional
detonations. As we have seen in Section \ref{sub:The-spectral-stability},
multi-dimensional instabilities can be dominant in the asymptotic
model, with the dispersion relation showing a maximum growth rate
for some nonzero transverse wave number $l$ (see Fig. \ref{fig:Spectrum-2d}).
In this section, we calculate the long-term dynamics of detonation
waves when two-dimensional effects are present. In particular, we
show that the asymptotic model retains the essential complexity required
to reproduce multi-dimensional cellular patterns. Solving (\ref{eq:asymptotic-model-inviscid-u}-\ref{eq:asymptotic-model-inviscid-lambda})
numerically turns out to be a non-trivial task and requires special
care. In the following subsection, we present a discussion of the
algorithm employed in this work.

\subsubsection{Numerical algorithm for the two-dimensional asymptotic system\label{sub:Numerical-algorithm}}

In order to appreciate the subtlety associated with (\ref{eq:asymptotic-model-inviscid-u}-\ref{eq:asymptotic-model-inviscid-lambda}),
we note that the equations comprise a nonlinear hyperbolic system
with one of its characteristic planes orthogonal to time. This means
that:
\begin{itemize}
\item The initial data are given on a characteristic surface, the $x-y$
plane. The absence of a time derivative in (\ref{eq:asymptotic-model-v}-\ref{eq:asymptotic-model-lambda})
requires the initial conditions to satisfy $v_{x}=u_{y}$, $\lambda_{x}=\omega\left(\lambda,u\right)$. 
\item Evolving in $\tau$ is a nonlocal procedure and, in the presence of
a shock, care has to be taken to avoid spurious numerical oscillations. 
\end{itemize}
Many existing numerical methods for solving (\ref{eq:UTSD-1}-\ref{eq:UTSD-2})
are based on a formal rewriting of the system as a single equation,
\begin{equation}
u_{xt}+\left(\frac{u^{2}}{2}\right)_{xx}+u_{yy}=0,\label{eq:UTSD-cross-differentiated}
\end{equation}
by cross-differentiation and substitution. Two concerns arise with
this approach. First, the validity of such a transformation is not
obvious when $u$ and $v$ are discontinuous functions. Second, in
the presence of chemical reactions, differentiation of (\ref{eq:asymptotic-model-inviscid-u})
with respect to $x$ produces a delta forcing at the ignition-temperature
locus due to the discontinuous nature of the reaction rate. Thus,
the techniques based on solving (\ref{eq:UTSD-cross-differentiated})
are inadequate for our purposes. 

Another common technique is to solve for the potential, $\phi$, that
satisfies 
\begin{equation}
\phi_{x\tau}+\left(\frac{\phi_{x}}{2}\right)_{x}^{2}+\phi_{yy}=0.
\end{equation}
Again, when chemical reactions are present, complications arise because
the reaction rate now becomes an exponential function of $\phi_{x}$
and the discretization errors in $\phi_{x}$ therefore exponentially
amplify, requiring very high-order methods for good accuracy. 

The method we employ here is based on a direct semi-implicit discretization
of (\ref{eq:asymptotic-model-u}-\ref{eq:asymptotic-model-v})\textbf{
}following some of the ideas found in\textbf{ }\cite{Hunter2000}.
In this method, all terms except for $v_{y}$ are treated explicitly.
By chosing second-order spatial and temporal discretizations, we obtain
an algorithm that is formally second order in time and space. The
general procedure is outlined here to explain our reasoning for the
choice of the algorithm. 

Assuming that the solution at time $\tau=\tau^{n}$ is known, we evolve
it to $\tau=\tau^{n+1}$ as follows:
\begin{enumerate}
\item First, we employ a semi-implicit time discretization, where the $v_{y}$
term is treated implicitly. The motivation for this comes from the
fact that some waves propagate infinitely fast in the $x-y$ plane.
In the simple case of a forward Euler time discretization, we obtain
\begin{eqnarray}
\frac{u^{n+1}-u^{n}}{\Delta\tau}+(F^{n}(u))_{x}+v_{y}^{n+1} & = & \lambda_{x}^{n},\label{eq:algorithm-semi-discrete-1}\\
v_{x}^{n+1} & = & u_{y}^{n+1},\label{eq:algorithm-semi-discrete-2}\\
\lambda_{x}^{n+1} & = & \omega\left(\lambda^{n+1},u^{n+1}\right),\label{eq:algorithm-semi-discrete-3}
\end{eqnarray}
where $F(u)=u^{2}/2$. A more quantitative reason for treating $v_{y}$
implicitly can be seen from the von Neumann stability analysis of
the linearized system without chemical reactions, wherein a fully
explicit scheme can be unstable (see Appendix \ref{sec:von-Neumann-stability-analysis}
for details).
\item We approximate the explicit terms $\lambda_{x}^{n}$ and $F_{x}^{n}$
using a shock-capturing scheme, e.g., finite volume or ENO/WENO algorithms
\cite{leveque2002finite}. 
\item Using (\ref{eq:algorithm-semi-discrete-2}), we write a forward difference
representation of $v_{i,j}^{n+1}$ in terms of $u_{i,j}^{n+1}$ and
$u_{i',j}^{n+1}$ for $i'>i$. For instance, a first-order forward
difference scheme can be used, i.e., 
\begin{equation}
v_{i,j}^{n+1}=v_{i+1,j}^{n+1}-\Delta x\left(u_{i,j}^{n+1}\right)_{y},\label{eq:algorithm-vx-discretization}
\end{equation}
where the $y$ derivative approximation is postponed until the next
step. The use of the forward difference here is a consequence of up-winding
the infinitely fast waves propagating from right to left. 
\item Approximate the $y$ derivatives, e.g., by centered differences, to
obtain the fully discrete scheme: 
\begin{eqnarray}
u_{i,j}^{n+1} & = & u_{i,j}^{n}+\Delta\tau\left[-\frac{1}{2}\left(\lambda_{x}\right)_{i,j}^{n}-\left(F_{x}\right)_{i,j}^{n}-\frac{1}{2\Delta y}\left(v_{i,j+1}^{n+1}-2v_{i,j}^{n+1}+v_{i,j-1}^{n+1}\right)\right],\label{eq:algorithm-u-ij}\\
v_{i,j}^{n+1} & = & v_{i+1}^{n+1}-\frac{\Delta x}{2\Delta y}\left(u_{i,j+1}^{n+1}-u_{i,j}^{n+1}+u_{i,j-1}^{n+1}\right).\label{eq:algorithm-v-ij}
\end{eqnarray}

\item In order to solve (\ref{eq:algorithm-u-ij}-\ref{eq:algorithm-v-ij}),
sweep from right to left, assuming the right boundary values of $u$
and $v$ are known at all times, in the following way:

\begin{enumerate}
\item For some fixed $i$, insert (\ref{eq:algorithm-v-ij}) into (\ref{eq:algorithm-u-ij})
and solve the linear system for the vector $u_{i,j}^{n+1}$.
\item With $u_{i,j}^{n+1}$ find $v_{i,j}^{n+1}$ using (\ref{eq:algorithm-u-ij}).
\item Repeat (a),(b) with $i=i-1$ until the left boundary is reached.
\end{enumerate}
\item Finally, compute $\lambda^{n+1}$ by solving (\ref{eq:algorithm-semi-discrete-3})
with a boundary condition on the right of the domain, which is given
by 
\begin{equation}
\lambda_{x}^{n+1}(x_{right},y)=0.
\end{equation}
Notice that (\ref{eq:algorithm-semi-discrete-3}) is actually an initial
value problem for $\lambda^{n+1}$ for fixed $y$ wherein $-x$ is
a time-like direction. It can be solved with any initial value solver
(e.g., a Runge-Kutta method) if desired.
\end{enumerate}
It may seem counterintuitive at first that the simplified model requires
a semi-implicit method while the reactive Euler equations can be solved
explicitly. The reason is that the asymptotic approximation is performed
in a limit where the reactive Euler equations themselves would have
to be treated implicitly. In order to understand this, we look at
the three waves present in the Euler equations. Since the weak heat
release approximation implies that the detonation velocity is nearly
acoustic, and an acoustic wave induces no flow behind it, we see that
the speed of the forward acoustic characteristic, given by $u+c-D$
in a frame moving with the wave, is actually an $O\left(\epsilon\right)$
quantity ($D\approx c$ and $u\approx\epsilon$). The entropy and
backward acoustic characteristics, on the other hand, have $O\left(1\right)$
speeds. In the asymptotic model, a slow time, $\tau$, is chosen so
that the dynamics happen on $O\left(1\right)$ time scales and therefore
some characteristics have speeds of $O\left(1/\epsilon\right)$ in
the slow time variable. When an explicit method is used to solve the
Euler equations in this limit, a typical CFL condition would require
a time step, $\Delta\tau\sim\epsilon\Delta x$, and as $\epsilon\to0$,
it becomes clear that the time step restriction becomes unattainable
and an implicit method is needed. 

The infinite characteristic velocity arises because the $x,y$ plane
is a characteristic surface for the equations. A simple rotation in
space-time \textquotedbl{}resolves\textquotedbl{}, to some extent,
this issue such that in the \textquotedbl{}rotated\textquotedbl{}
space-time the system behaves as a standard system of conservation
laws. The trade-off for recovering a classical hyperbolic system with
finite wave speeds is that one must now consider a grid with moving
boundaries. This idea can be exploited to produce fully explicit schemes
as shown in \cite{tabak1994focusing} for the case without reactions.

\subsubsection{Two-dimensional cellular detonations\label{sub:Two-dimensional-cellular-detonations}}

Using the algorithm described in the previous section, we solve (\ref{eq:asymptotic-model-inviscid-u}-\ref{eq:asymptotic-model-inviscid-lambda})
in a frame moving with constant speed, $D_{0}=1$. All simulations
are initialized with the one-dimensional ZND solutions obtained in
Section \ref{sub:Travelling-wave-solutions}. We then investigate,
in the context of the asymptotic equations: 1) the effect of the width
of the channel on nonlinear stability properties of the travelling
wave solutions; 2) the effect of the periodic boundary conditions;
and 3) the effect of increasing heat relase on the wave dynamics. 

\begin{figure}[H]
\label{fig:detonation-on-channel-vary-q}\subfloat[$L_{y}=2$]{\includegraphics[width=3in,height=2in]{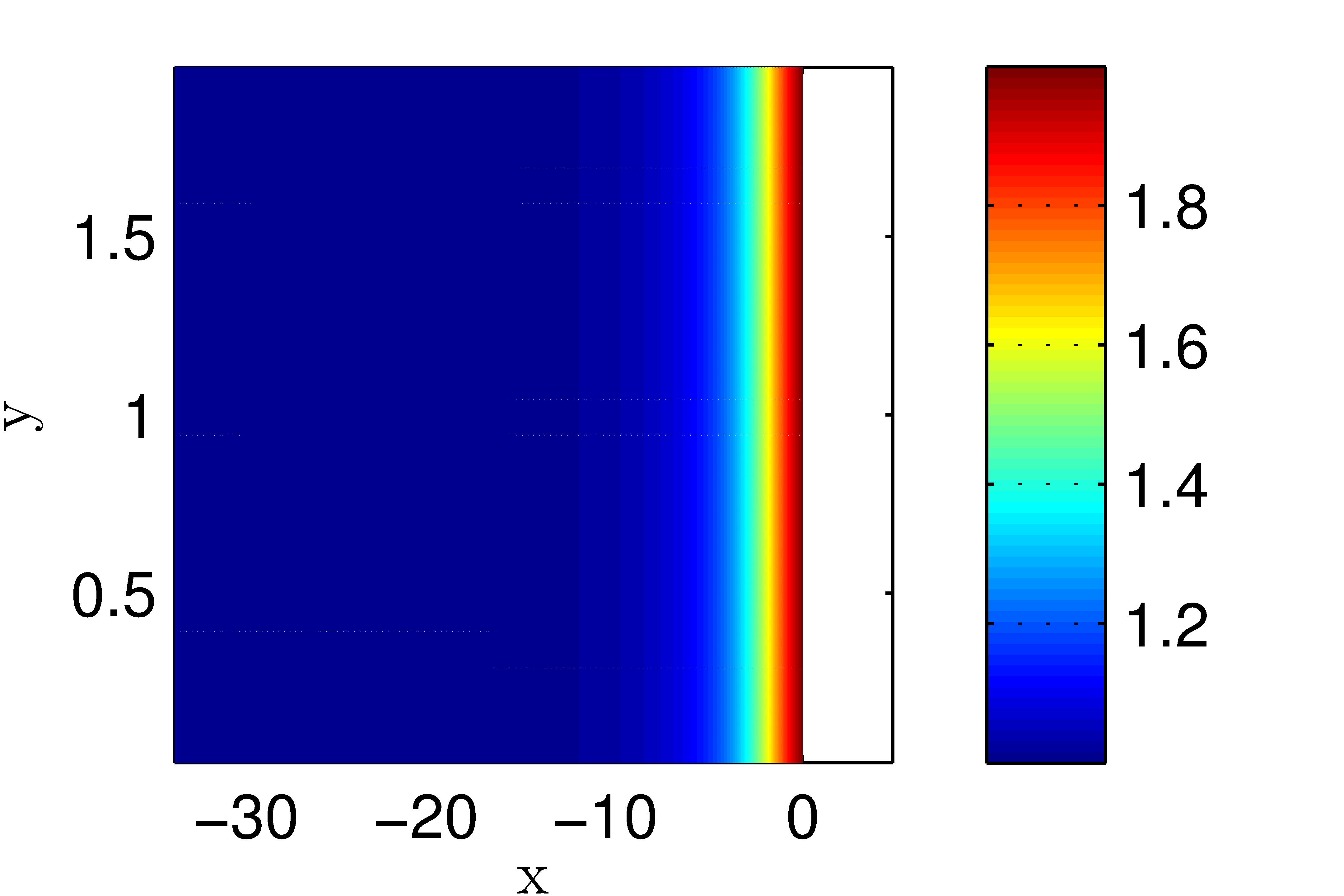}}\hfill\subfloat[$L_{y}=10$]{\includegraphics[width=3in,height=2in]{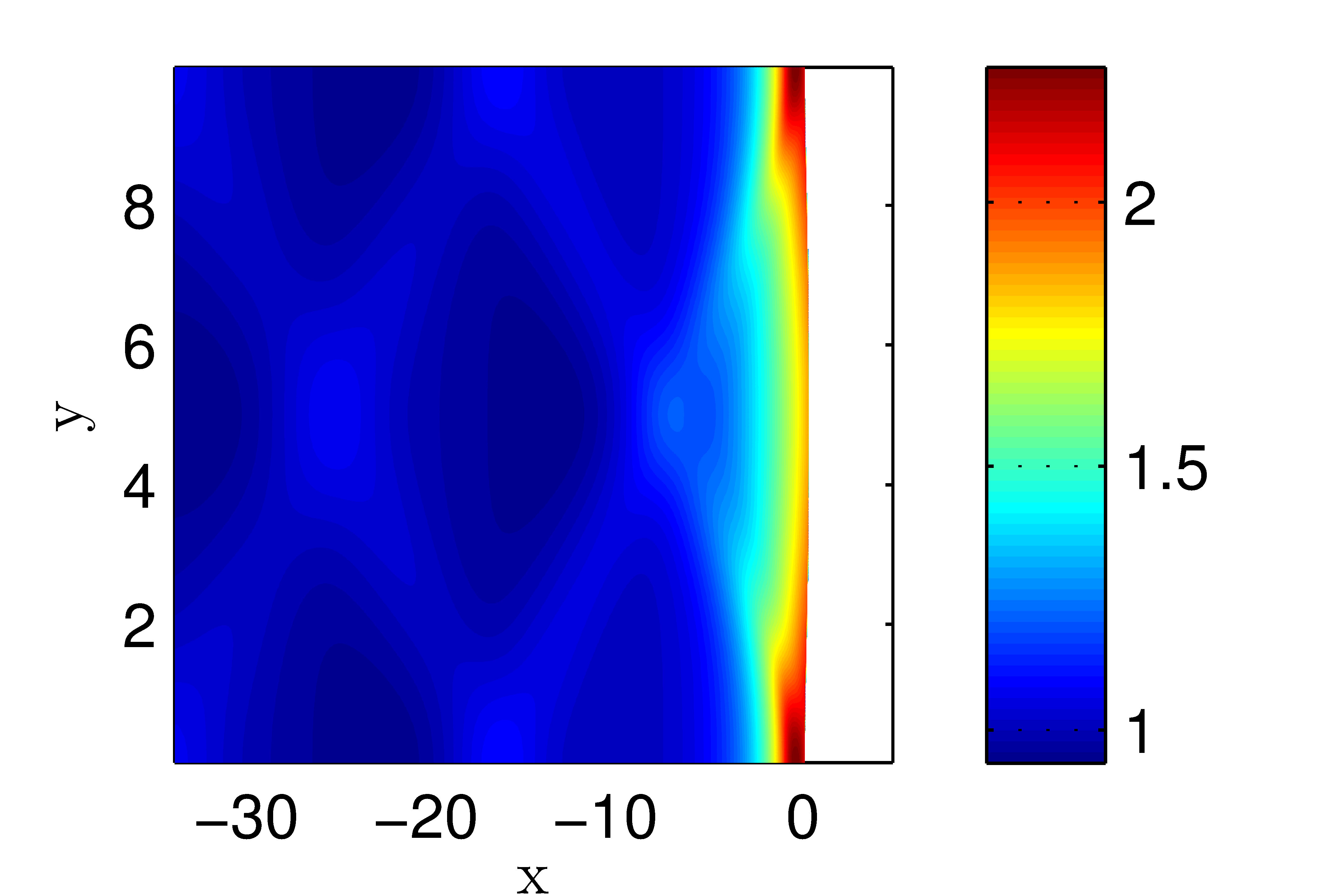}}\\
\subfloat[$L_{y}=50$]{\includegraphics[width=3in,height=2in]{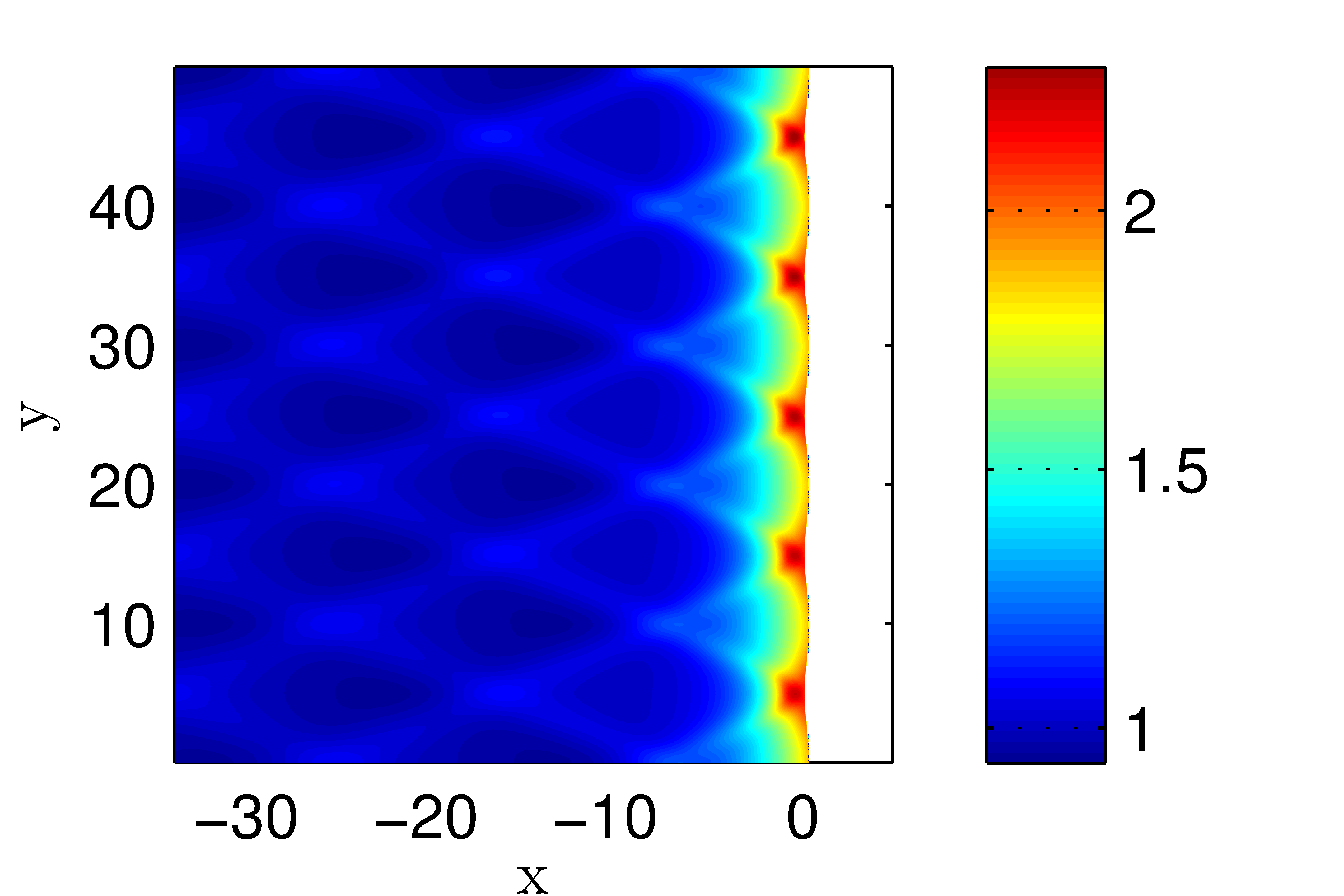}}\hfill\subfloat[$L_{y}=100$]{\includegraphics[width=3in,height=2in]{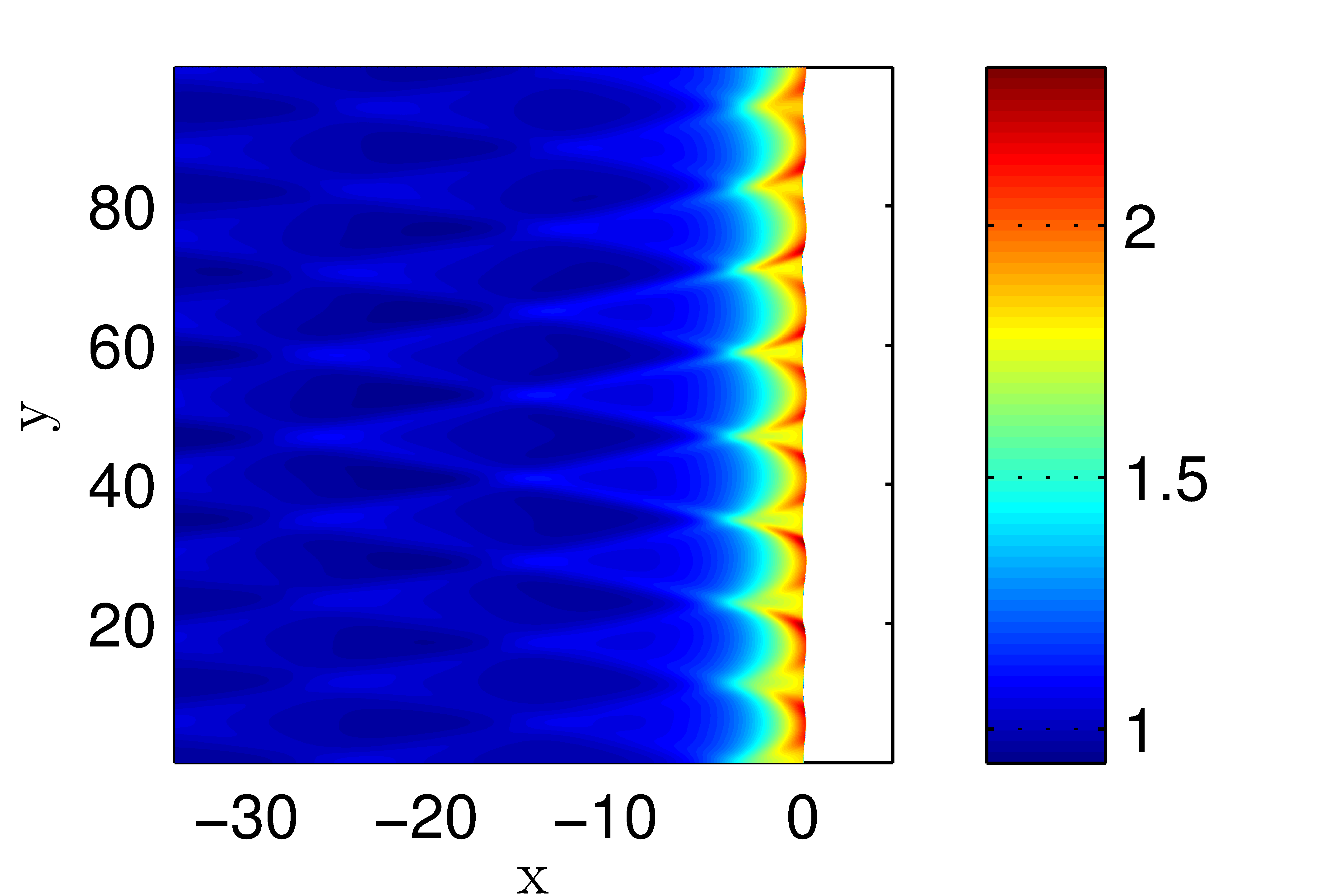}}

\caption{\label{fig:detonation-in-channel-vary-size}Dynamics of the asymptotic
model in channels of different widths. The plots show the asymptotic
variable $u$. The parameters are $q=1.7$, $\theta=1.65$, as in
Fig. \ref{fig:Spectrum-2d}. The white region corresponds to the ambient
state ahead of the wave, with $u=0$.}
\end{figure}

In Fig. \ref{fig:detonation-in-channel-vary-size}, we show the profile
of $u$ at $\tau\approx500$ for varying widths of the domain, $L_{y}$,
in the $y$ direction. The $y$ boundaries are modelled as rigid walls,
appropriate for a detonation in a two-dimensional channel (i.e., a
channel with negligible depth). The parameters are fixed at $q=1.7$
and $\theta=1.65$, for which the linear stability calculation shows
that the ZND wave is unstable only to two-dimensional perturbations.
In a channel of width $L_{y}=2$, the ZND wave remains stable, as
seen in Fig. \ref{fig:detonation-in-channel-vary-size}(a), as the
spacing is too narrow for the two-dimensional instability to develop.
Increasing the width of the channel, however, allows for a number
of transverse modes to be excited%
\footnote{Recall that for a channel with finite width, $L_{y}$, the allowed
transverse wave numbers, $l$, are given by $l=\pi n/L_{y}$, where
$n\in\mathbb{Z}$. Therefore, only a discrete set of modes can be
excited, and larger $L_{y}$ typically allows for more unstable modes
to appear. %
}, leading to the formation of a multi-dimensional cellular pattern
as shown in Figs. \ref{fig:detonation-in-channel-vary-size}(c,d).
The cell size appears to remain around $10$ in all Figs. \ref{fig:detonation-in-channel-vary-size}(b-d).

\begin{figure}[H]
\centering{}\includegraphics[width=6in,height=2in]{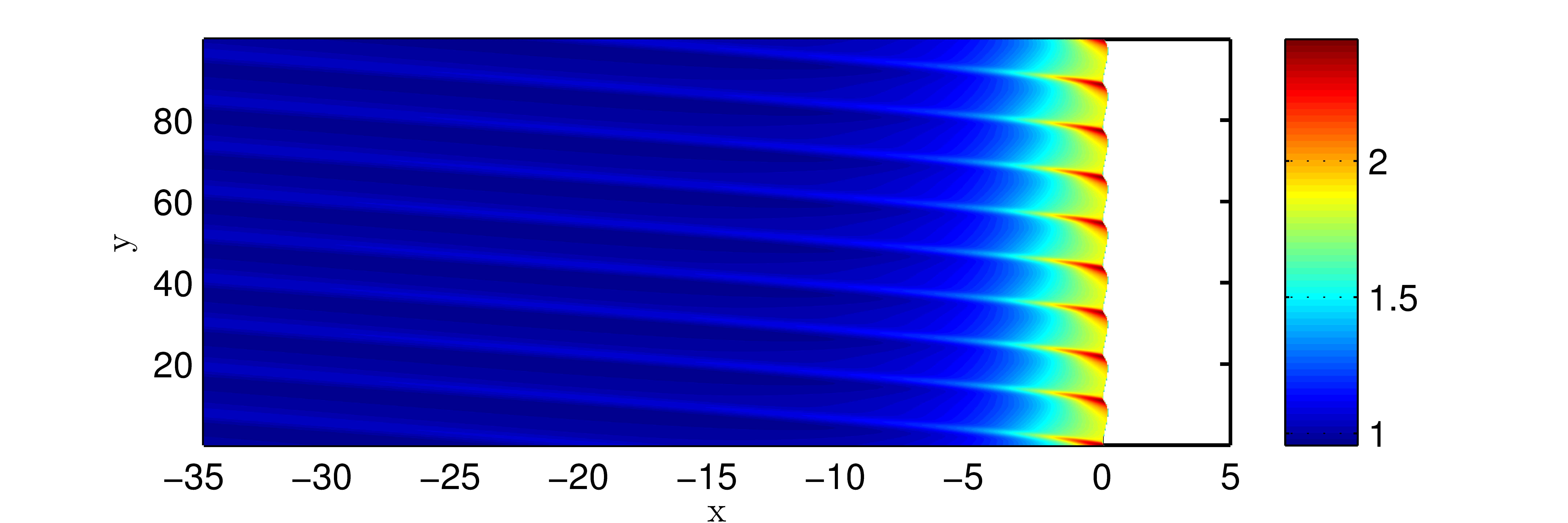}\caption{\label{fig:Spinning-detonation-in-tube}Detonation in a channel of
width $L=100$ with periodic boundary conditions. Parameters are $q=1.7$
and $\theta=1.65$. }
\end{figure}

When we replace the solid wall boundary conditions with periodic conditions,
we observe a two-dimensional version of a spinning detonation. There
is only one family of transverse shock waves that all propagate in
the same direction. Such a wave can be imagined to form in a narrow
gap between two concentric cylindrical tubes, as in a rotating detonation
engine \cite{VMT63}. A snapshot of the solution field, $u$, is shown
in Fig. \ref{fig:Spinning-detonation-in-tube}, where relatively strong
transversely propagating shocks can be seen. 


\begin{figure}[H]
\subfloat[$q=1.7$]{\includegraphics[width=3in,height=2in]{Figures/2d-asymptotic-tube-L-100-q-1_7}}\hfill\subfloat[$q=1.8$]{\includegraphics[width=3in,height=2in]{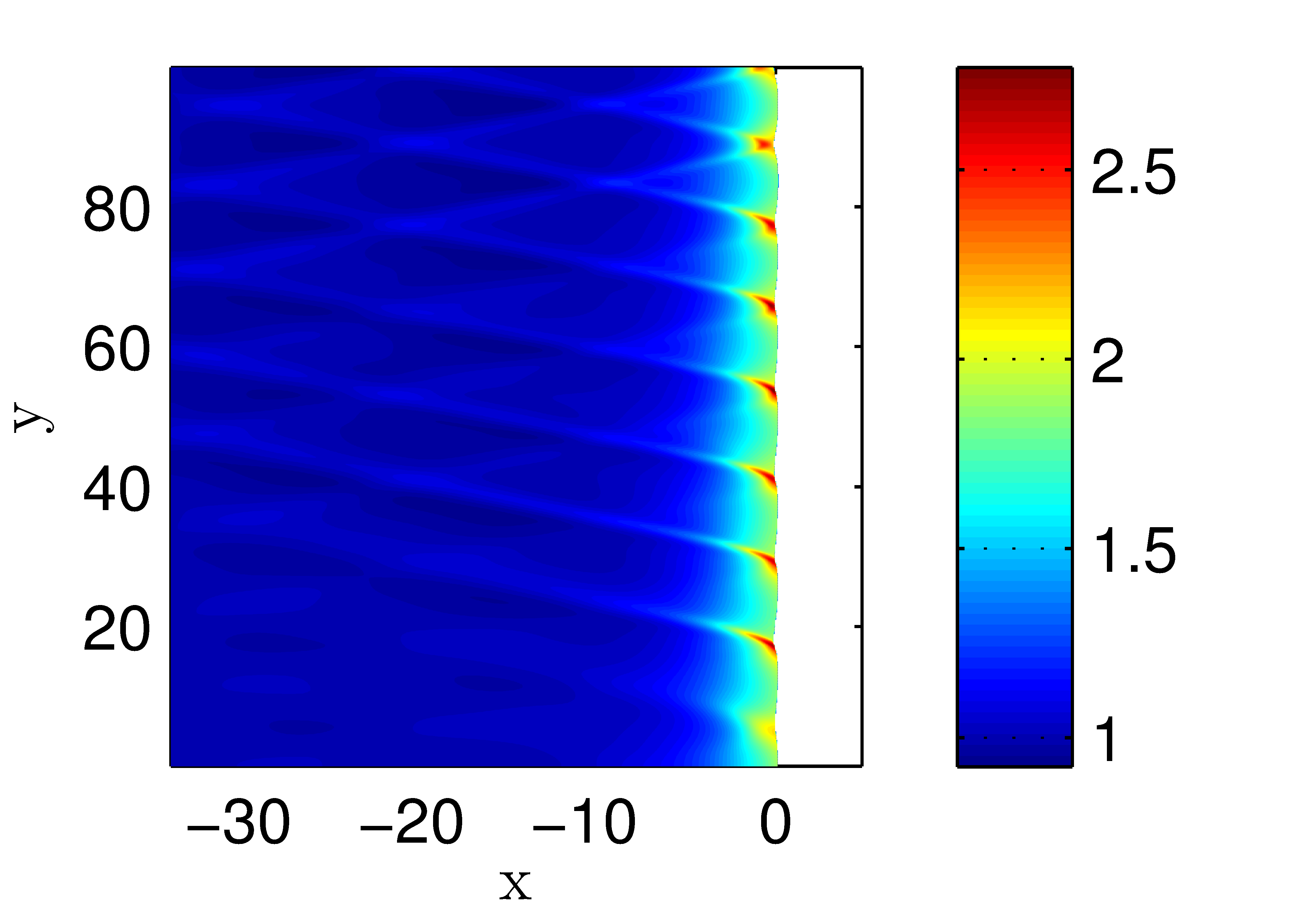}}\\
\subfloat[$q=1.9$]{\includegraphics[width=3in,height=2in]{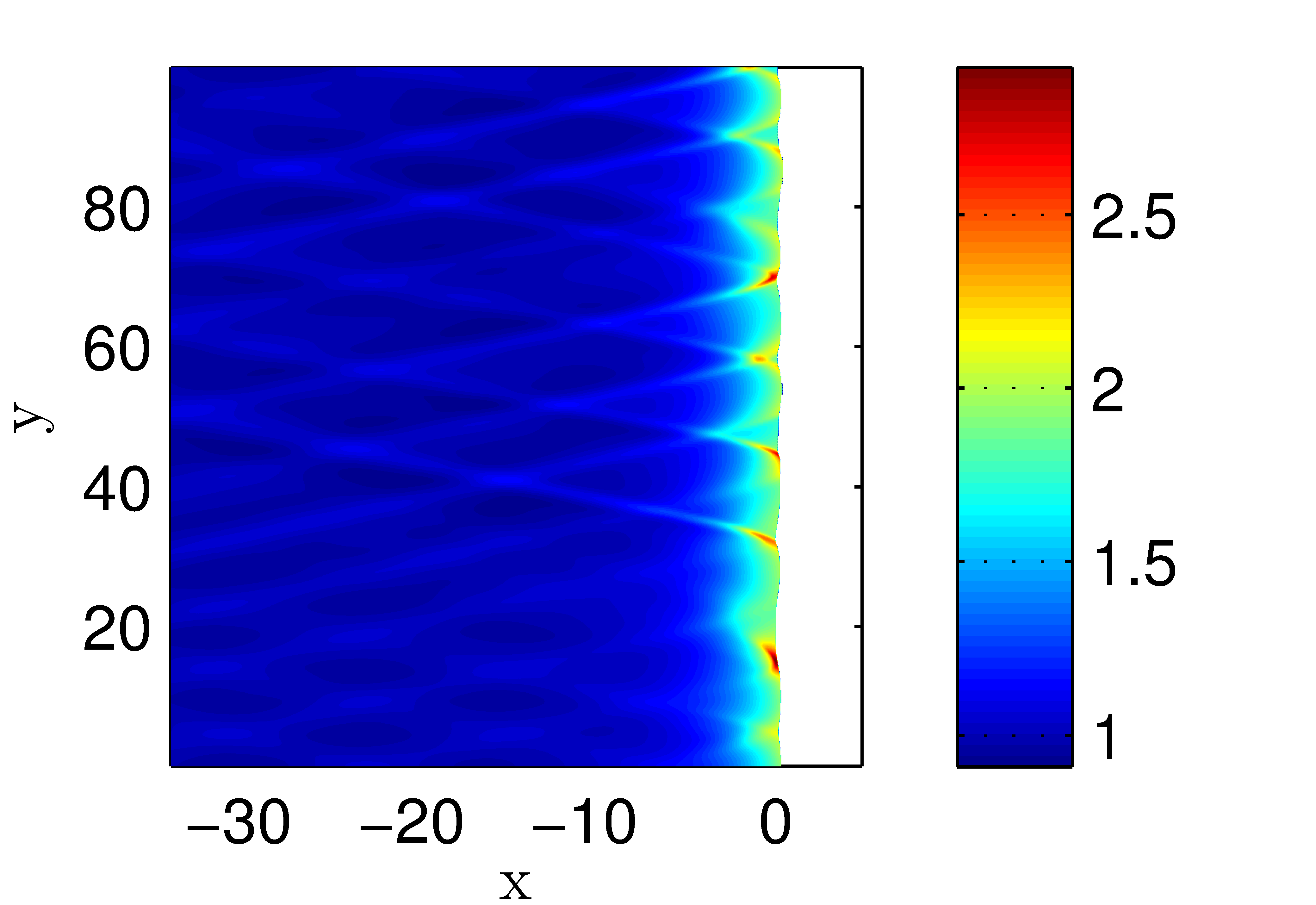}}\hfill\subfloat[$q=2.5$]{\includegraphics[width=3in,height=2in]{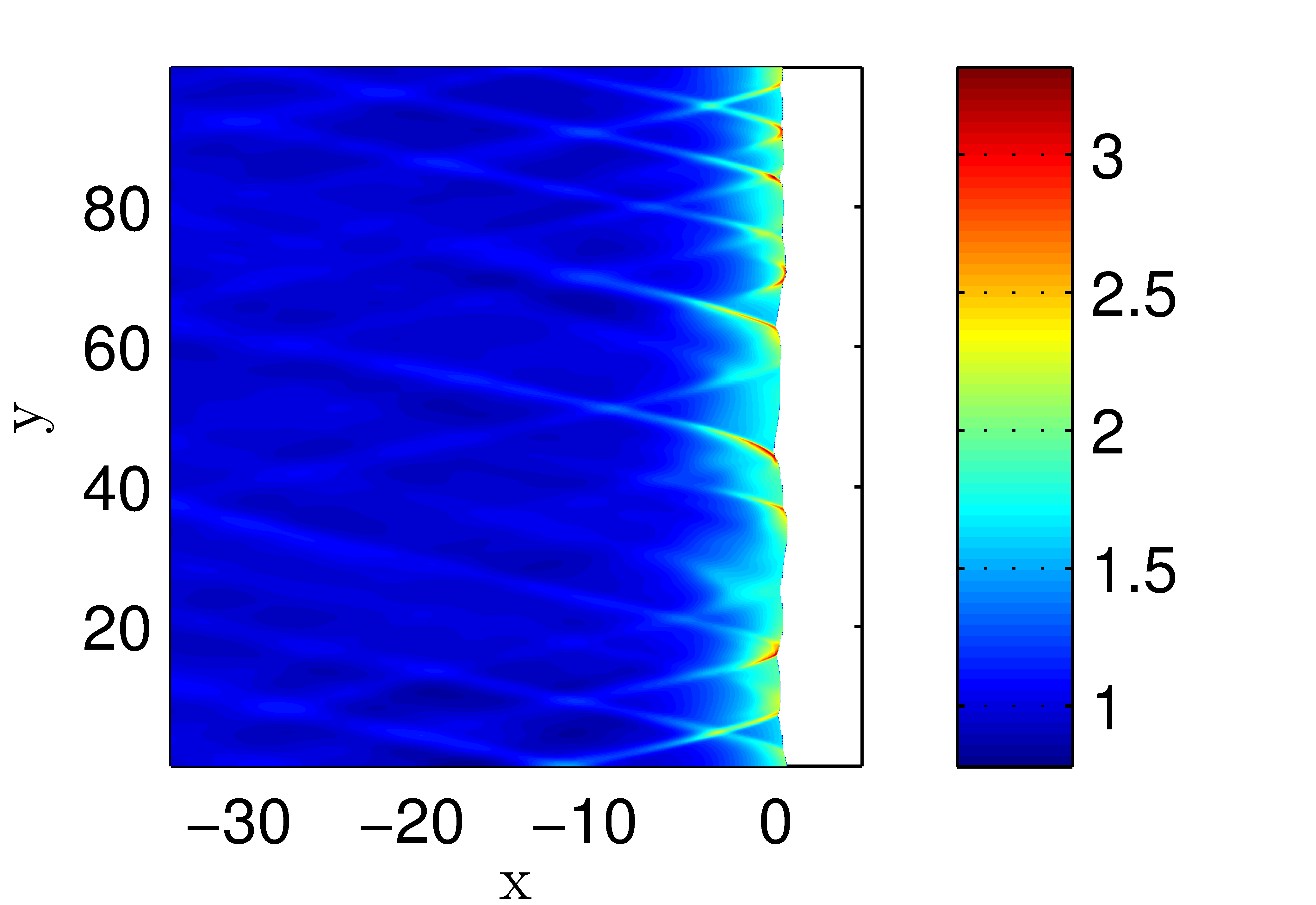}}
\caption{\label{fig:detonation-in-channel-vary-q}Dynamics of the asymptotic
model for varying heat release at a fixed width $L=100$ and activation
energy $\theta=1.65$.}
\end{figure}

\begin{figure}[H]
\subfloat[Density field, $\rho=1+\epsilon\sqrt{q}u$, as predicted by the asymptotic
model.]{\includegraphics[width=6in,height=2in]{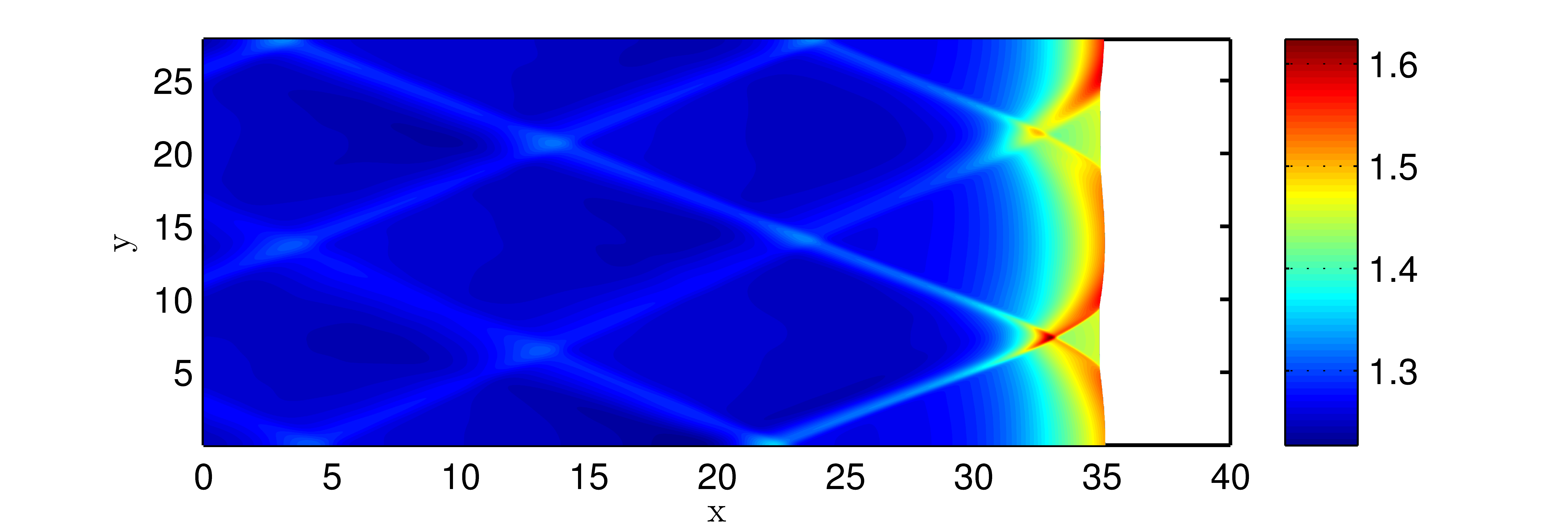}}\\
\subfloat[Density field computed from the reactive Euler equations.]{\includegraphics[width=6.3in,height=2in]{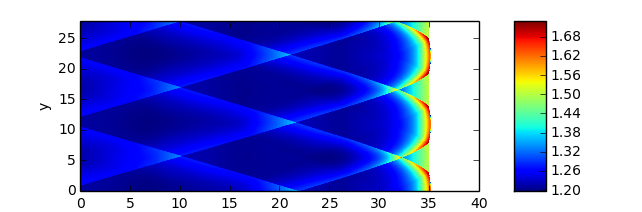}}\caption{\label{fig:Nonlinear-dynamics-2d}Comparison between asymptotic and
full solutions of the reactive Euler equations for a cellular detonation
in a channel. }
\end{figure}

In Fig. \ref{fig:detonation-in-channel-vary-q}, we show the effect
of the heat release on the solution structure. We fix $\theta=1.7$
and increase the value of $q$ from $1.7$ to $2.5$. Consistent with
the behaviour of detonations in the reactive Euler equations, regular
cells are observed at small $q$, as in Figs. \ref{fig:detonation-in-channel-vary-q}(a,b),
but with increasing $q$, the structure of the detonation front becomes
more complex with the formation of irregular cells as in Fig. \ref{fig:detonation-in-channel-vary-q}(d). 
\\
\indent All of the previous results show that, at least at a qualitative level,
the asymptotic model captures many important characteristics of multi-dimensional
detonations. Next, we investigate how \emph{quantitatively} close
the predictions are to the solutions of the reactive Euler equations.
The numerical simulations of the reactive Euler equations were carried
out using PyCLAW \cite{pyclaw-sisc}, which provides, among other
things, a Python wrapper for the classic routines in CLAWPACK. The
software requires the user to provide a Riemann solver. We use a Roe-linearized
Riemann solver with a Harten-Hyman entropy fix \cite{leveque2002finite}.
The classic package of PyCLAW employs a second-order finite volume
algorithm with a fractional step method for the time evolution \cite{leveque2002finite}.
For the numerical simulations, the reactive Euler equations were non-dimensionalized
in the conventional form by the ambient state with velocities scaled
by $u_{a}=\sqrt{p_{a}/\rho_{a}},$ spatial variables scaled by the
half-reaction length, $x_{1/2}$, and time by $u_{a}/x_{1/2}$. 
\\
\indent In Fig. \ref{fig:Nonlinear-dynamics-2d}, we show a comparison between
the asymptotic solutions and the solutions of the reactive Euler equations.
The parameters chosen are the same as in Fig. \ref{fig:Spectrum-2d-euler-vs-asymptotic}(b),
i.e., $\gamma=1.2$, $Q=0.4$ and $E=50$, which in the asymptotic
variables are given by $q=2.4$ and $\theta=50/36\approx1.389$. We
start both simulations with the ZND solution and solve for a time
interval large enough such that instabilities have already fully developed.
Notice that since the asymptotic model uses the slow time variable,
$\tau$, we only need to solve it for a relatively short time interval,
$\tau\approx100$. The reactive Euler system, on the other hand, was
non-dimensionalized in the conventional way using the regular time
variable, $t$, and thus we must solve the system up to $t\approx\tau/\epsilon=600$.
We perform all necessary scaling conversions between the variables
so that the lengths and amplitudes shown in Fig. \ref{fig:Nonlinear-dynamics-2d}
are consistent. The length of the $y$ domain is fixed in the asymptotic
model to be $20$, which corresponds to a length $27.83$ in the regular
dimensionless $y$, while the $x$ domain is of length $40$. 

We see from Fig. \ref{fig:Nonlinear-dynamics-2d} that the asymptotic
model captures the salient features of multi-dimensional detonations
with good quantitative agreement. The characteristic scales of the
detonation cells are seen to be close. A small difference in the transverse
propagation velocity of the triple points is apparent, in the asymptotic
case the speed being higher. The transverse shocks in the asymptotic
solutions far from the lead shock are seen to be smoother than in
the full solutions, which is due to a larger numerical diffusion in
the algorithm for the asymptotic model.

\begin{figure}[H]
\begin{centering}
\includegraphics[width=4in,height=3in]{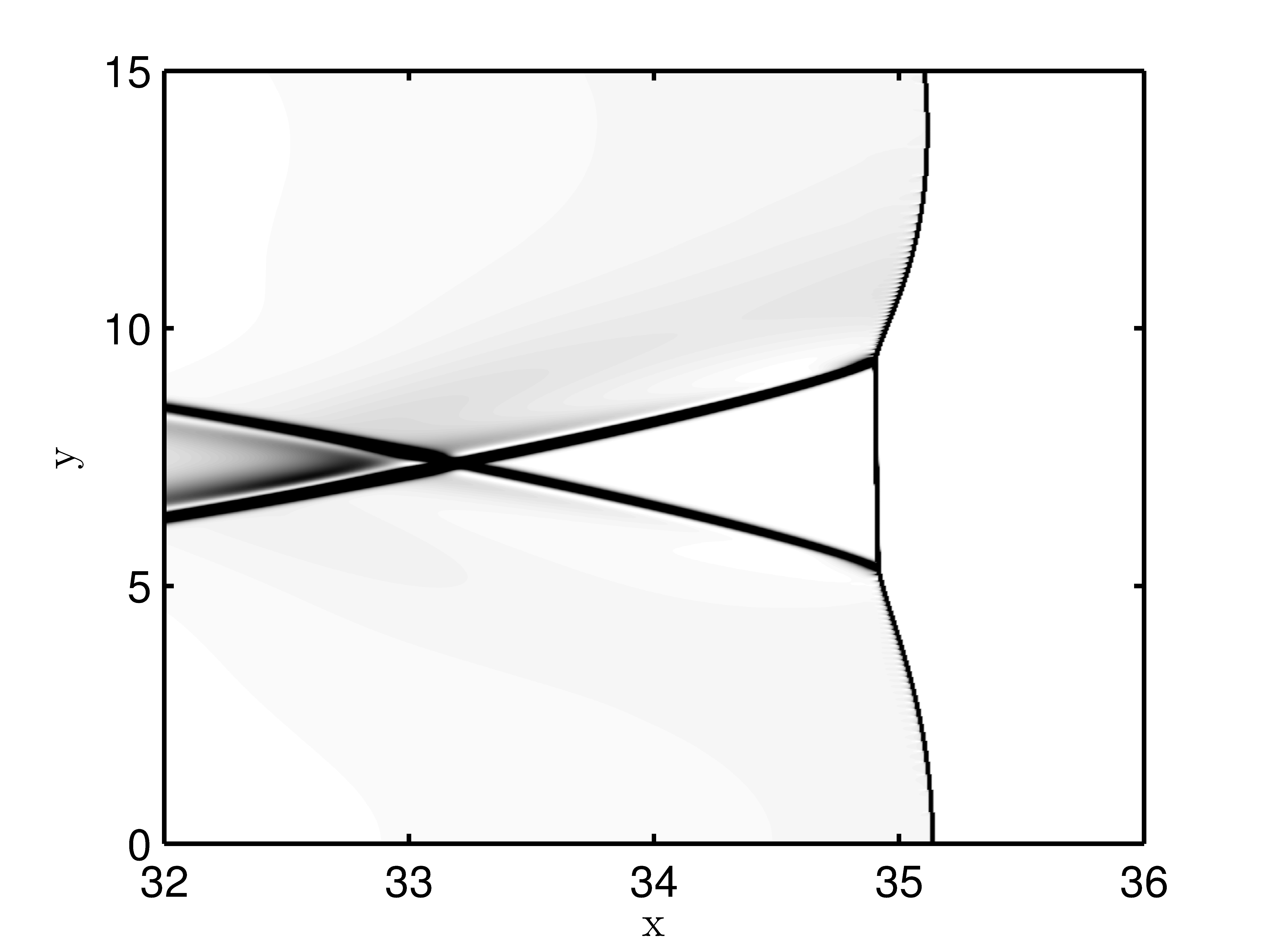}
\par\end{centering}

\caption{\label{fig:Triple-point-like-structures} Triple-point-like structures
in the asymptotic solution of the weakly nonlinear detonation as shown
by the magnitude of the density gradient.}
\end{figure}

Finally, we remark that weakly nonlinear detonations operate in the
same fluid dynamic regime as weak shock wave focusing or weak shock
reflection at near grazing incidence. As pointed out by von Neumann
\cite{vonNewmann1943oblique}, in this limit the \textquotedbl{}classical\textquotedbl{}
Mach triple point shock structure is impossible. Yet, both experiments
and numerical calculations \cite{glaz1985numerical} consistently
exhibit triple shock structures in this regime. This apparent contradiction
is known as the \textquotedbl{}von Neumann paradox\textquotedbl{}.
Although the structures behind the lead shock in our calculations
resemble the triple-points observed in detonations (Fig. (\ref{fig:Triple-point-like-structures})),
a simple algebraic argument can show that the asymptotic equations
do not admit ``classical'' triple points, where three shocks separated
by smooth flows meet \cite{tabak1994focusing,Hunter2000}. Thus, much
like in the problem of weak shock reflection, the observed cells in
the weak heat release detonations considered in this work present
yet another instance of the ``von-Neumann paradox''. Here, we make
no attempt to address this problem and simply note that triple-point-like
structures appear in our numerical simulations behind the lead shock;
whether they are true discontinuities, sharp waves or some singularities
remains an open problem. 

\section{Discussion and conclusions\label{sec:Discussion-and-conclusions}}

In this work, we develop an asymptotic theory of multi-dimensional
detonations within the framework of the compressible Navier-Stokes
equations for a perfect gas reacting with a one-step heat-release
law. The main outcome is a reduced model that consists, in two spatial
dimensions, of a forced Burgers-like equation coupled with a heat
release equation and an equation that enforces zero vorticity. After
specializing the model to the case without dissipative effects, we
have analyzed it in detail and have shown that it captures many of
the dynamical features of detonations in reactive Euler equations.
These include: 1) existence of steady-state travelling waves (i.e.,
ZND solutions), 2) linear instability of the ZND solutions, 3) existence
of a period-doubling bifurcation road to chaos in the one-dimensional
case, and 4) onset of cellular detonations in the two-dimensional
case. Our theory considers weakly nonlinear detonations in the distinguished
limit of small heat release, large activation energy and small $\gamma-1$.
In this limit, the dynamics occur on time scales that are long compared
with the scale of a characteristic chemical reaction time. Our reduced
evolution system builds on the weakly nonlinear theory in \cite{RosalesMajda:1983ly,rosales1989diffraction}
by adding the Newtonian approximation to it. In the one-dimensional
and non-dissipative special case, our theory contains the model in
\cite{clavin2002dynamics}.

As a consequence of the $\gamma-1=O(\epsilon)$ assumption, the temperature
expansion in our analysis starts with an $O\left(\epsilon^{2}\right)$
correction to the leading-order term, as opposed to $O\left(\epsilon\right)$
correction in \cite{rosales1989diffraction,RosalesMajda:1983ly}.
This is precisely what allows us to escape the fact that in \cite{rosales1989diffraction,RosalesMajda:1983ly}
there must be leading-order corrections to temperature, velocity and
density that behave exactly the same way. The present scaling highlights
an important difference between a reactive and a non-reactive shock,
namely, that the temperature in a reactive shock can increase at some
distance from the shock front because of heat release. This increase
in the post-shock temperature means that the energy release can possess
a maximum inside the reaction zone. As we previously elucidated with
a reactive Burgers model \cite{kasimov2013model,FariaKasimovRosales-SIAM2014},
the presence of such an internal maximum appears to be a key factor
for the mechanism that generates resonances and subsequent amplification
of the waves in the reaction zone. 

Note that all one-dimensional weakly nonlinear asymptotic models as
well as the \emph{ad hoc} models devised to capture detonation dynamics
\cite{Fickett:1979ys,Majda:1980zr,Radulescu:2011fk,kasimov2013model}
are extensions of the Burgers equation modified by adding a reaction
forcing term, with an extra equation for the reaction progress variable.
The reason is that the dynamics of any one-dimensional genuinely nonlinear
hyperbolic system reduces to that of an inviscid Burgers equation
in the weakly nonlinear limit \cite{rosales1991introduction}.

While the reduced weakly nonlinear asymptotic theory has clear predictive
power, as demonstrated in this work, recognizing its limitations is
important and helps highlight open problems that require further investigation.
For example, at parameter values (especially of the heat release)
corresponding to strong cellular and pulsating detonations typical
in experiments and numerical simulations, the asymptotic predictions
are not sufficiently accurate when compared to the reactive Euler
equations. While it is not clear how to extend the present approach
directly to strongly nonlinear detonations, the ability of the reduced
system obtained in this work to capture essential dynamical characteristics
of unsteady and multi-dimensional detonations serves as a strong indication
that a theory of comparable simplicity may be possible for strongly
nonlinear detonations. 

Recent work \cite{barker2013viscous} on the role of dissipation in
the compressible reactive Navier-Stokes system in one spatial dimension
indicates that dissipation can have non-trivial effects on the stability
of detonation waves. Elucidation of the role played by the transport
effects in the asymptotic model is therefore of interest and we believe
that the model retaining these effects, as in (\ref{eq:asymptotic-model-u}-\ref{eq:asymptotic-model-lambda}),
should be investigated further.

A problem that merits exploration, in our opinion, is that of the
evolution of a small-amplitude, localized initial perturbation to
a detonation wave, considered in the same weakly nonlinear asymptotic
regime as in the present work. All of the underlying assumptions in
the theory remain valid for this different initial-value problem,
which is relevant to the problem of detonation initiation.

From a mathematical point of view, the reduced system obtained in
this work and its connection with the issue of the non-existence of
triple-point shock structures for the Zabolotskaya-Khokhlov equation
pose some puzzling and challenging problems.

Finally, recall that the qualitative models introduced by Fickett
and Majda do not contain instabilities with the rate functions used
in prior work with these models (e.g., see \cite{humpherys2013stability}).
However, using the rate function derived in this paper (and, for that
matter, the two-step rate function used in \cite{Radulescu:2011fk})
can be shown to lead to the same complexity of the solutions as occurs
in the reactive Euler equations. A modified Fickett's analog can be
written as 
\begin{alignat}{1}
u_{t}+uu_{x}= & -\frac{1}{2}\lambda_{x},\label{eq:new-Fickett-Majda-1}\\
\lambda_{t}= & k\left(1-\lambda\right)\exp\left(\alpha u+\beta\lambda\right),\label{eq:new-Fickett-Majda-2}
\end{alignat}
with the ``activation energy'' and ``heat release'' parameters,
$\alpha$ and $\beta$, respectively. Clearly, one can treat more
general right-hand sides for the rate function in (\ref{eq:new-Fickett-Majda-2}),
such as $\omega=\psi\left(\lambda\right)\phi\left(u\right)$, as long
as $\psi$ and $\phi$ retain the same qualitative properties as their
corresponding expressions in (\ref{eq:new-Fickett-Majda-2}). This
modification, even though simple, reflects the physics of the phenomenon
responsible for the dynamics of pulsating detonations by allowing
a local maximum in the reaction rate to exist behind the precursor
shock in the detonation wave. Mathematical study of the modified analog
model is of interest from the point of view of the theory of hyperbolic
balance laws and the dynamics of their solutions.

\section*{Acknowledgements}

L.F. and A.K. gratefully acknowledge research support by King Abdullah
University of Science and Technology (KAUST). The research by R. R.
Rosales was partially supported by NSF grants DMS-1007967, DMS-1115278,
DMS-1318942, and by KAUST during his research visit in November 2013.
L. F. would like to thank Slava Korneev and David Ketcheson  for their
help with numerical computations.

\appendix

\section{\label{sec:von-Neumann-stability-analysis}von Neumann stability
analysis of a simple fully explicit scheme}

In order to motivate the use of a semi-implicit scheme to solve (\ref{eq:asymptotic-model-u}-\ref{eq:asymptotic-model-lambda}),
here we perform a von-Neumann stability analysis of a \textquotedbl{}natural/reasonable\textquotedbl{}
explicit scheme and show that it leads to instabilities. Since such
an analysis requires a constant coefficient linear system, we consider
here the linearized, unsteady, transonic small disturbance equations
\begin{eqnarray}
u_{t}+u_{x}+v_{y} & = & 0,\label{eq:linearized-UTSD-1}\\
v_{x} & = & u_{y}.\label{eq:linearized-UTSD-2}
\end{eqnarray}
We discretize these equations using forward finite differences in
$t$ and $x$, and centered finite differences in $y$. This leads
to the scheme

\begin{eqnarray}
U_{i,j}^{n+1} & = & U_{i,j}^{n}-\Delta t\left(\frac{U_{i,j}^{n}-U_{i-1,j}^{n}}{\Delta x}+\frac{V_{i,j+1}^{n}-V_{i,j-1}^{n}}{2\Delta y}\right),\\
V_{i,j}^{n} & = & V_{i+1}^{n}-\Delta x\left(\frac{U_{i,j+1}^{n}-U_{i,j-1}^{n}}{2\Delta y}\right).
\end{eqnarray}
Then we compute the (periodic) discrete eigenfunctions for the scheme
using the ansatz 
\begin{eqnarray}
U_{p,q}^{n} & = & AG^{n}e^{i\left(kx_{p}+ly_{q}\right)},\\
V_{p,q}^{n} & = & BG^{n}e^{i\left(kx_{p}+ly_{q}\right),}
\end{eqnarray}
where $k$ and $l$ are the discrete wave numbers, $G$ is the growth
factor, $A$ and $B$ are constants, $x_{p}=p\Delta x$, $y_{q}=q\Delta y$,
and 
\begin{eqnarray}
U_{p,q}^{n} & = & U(x_{q},y_{p},t^{n}),\\
V_{p,q}^{n} & = & V(x_{q},y_{p},t^{n}).
\end{eqnarray}

In the standard fashion of the von Neumann stability analysis, this
leads to an eigenvalue problem for the vector with components $A$
and $B$, with eigenvalue $G$. Solving this problem yields
\begin{eqnarray}
G & = & 1-\frac{\Delta t}{\Delta x}\frac{1-e^{-ik\Delta x}}{\Delta x}-\frac{\Delta t\Delta x}{\Delta y^{2}}\frac{\sin^{2}(l\Delta y)}{\left(1-e^{ik\Delta x}\right)}.\label{eq:von-Neumann-growth-factor}
\end{eqnarray}

Note that, in (\ref{eq:von-Neumann-growth-factor}), the term
\begin{equation}
\frac{\Delta t\Delta x}{\Delta y^{2}}\frac{\sin^{2}(l\Delta y)}{\left(1-e^{ik\Delta x}\right)}
\end{equation}
can be traced back to the explicit treatment of $v_{y}$. This term
causes instability, since it can become arbitrarily large for $k\Delta x$
small and $\sin^{2}\left(l\Delta y\right)$ away from zero, independently
of the size of $\Delta t$. Hence, the scheme is unstable. Instabilities
like this one are inevitable in explicit finite difference schemes,
regardless of the choice. The reason is that the wave speeds in the
$y$-direction are unbounded. No explicit scheme can hence satisfy
the CFL (Courant-Friedrichs-Lewy) condition.\bibliographystyle{plain}


\end{document}